\documentclass[apj,twocolumn,twocolappendix,numberedappendix]{openjournal}
\usepackage{amsmath}
\usepackage{booktabs}
\usepackage{multirow}
\usepackage{color}
\usepackage{soul}
\usepackage{booktabs}
\usepackage{threeparttable}

\usepackage{siunitx}
\usepackage{rotating}
\usepackage{graphicx}
\usepackage{appendix}
 
\usepackage[dvipsnames]{xcolor} 
 
\usepackage[breaklinks,colorlinks,citecolor=blue,urlcolor=blue]{hyperref}
\usepackage{xurl}

\newcommand{\Runit}{\mathrm{yr^{-1}\,Gpc^{-3}}\,h_{70}^{3}}

\defcitealias{Desai2024}{D24}
 
\renewcommand{\arraystretch}{1.2}
\usepackage{listings}
\usepackage{color}
\definecolor{dkgreen}{rgb}{0,0.6,0}
\definecolor{gray}{rgb}{0.5,0.5,0.5}
\definecolor{mauve}{rgb}{0.58,0,0.82}
\definecolor{golden}{rgb}{0.86,0.65,0.01}
\usepackage{orcidlink}

\begin{document}

\shorttitle{ASAS-SN Rates V: TDEs}
\shortauthors{Desai, Hinkle et al.}

\title{\textbf{ASAS-SN R\lowercase{ates} V: T\lowercase{idal} D\lowercase{isruption} E\lowercase{vent} V\lowercase{olumetric} R\lowercase{ates} \lowercase{and} L\lowercase{uminosity} F\lowercase{unctions} \lowercase{by} \\
O\lowercase{ptical} S\lowercase{pectroscopic} T\lowercase{ype}, X-\lowercase{ray} L\lowercase{uminosity}, \lowercase{and} H\lowercase{ost}-G\lowercase{alaxy} P\lowercase{roperties}}}

\author{\vspace{-1.4cm}
        Dhvanil~D.~Desai$^{\dagger}$\,\orcidlink{0000-0002-2164-859X}$^{1}$}
\author{Jason~T.~Hinkle$^{\dagger}$\,\orcidlink{0000-0001-9668-2920}$^{2,3,*}$}
\author{Benjamin~J.~Shappee\,\orcidlink{0000-0003-4631-1149}$^{1}$}
\author{Christopher~S.~Kochanek\,\orcidlink{0000-0001-6017-2961}$^{4,5}$}
\author{K.~Z.~Stanek\,\orcidlink{0009-0001-1470-8400}$^{4,5}$}
\author{Margaret~Shepherd\,\orcidlink{0009-0005-1158-1896}$^{2}$}
\author{Yashasvi~Moon\,\orcidlink{0009-0004-0436-0932}$^{2,3}$}
\author{John~F.~Beacom\,\orcidlink{0000-0002-0005-2631}$^{6,4,5}$}
\author{Subo~Dong\,\orcidlink{0000-0002-1027-0990}$^{7,8,9}$}
\author{Jose~L.~Prieto\,\orcidlink{0000-0003-1072-2712}$^{10}$}
\author{Todd~A.~Thompson\,\orcidlink{0000-0003-2377-9574}$^{4,6,5}$}
\author{Katie~Auchettl\,\orcidlink{0000-0002-4449-9152}$^{11}$}
\author{K.~Decker~French\,\orcidlink{0000-0002-4235-7337}$^{2}$}
\author{Willem~B.~Hoogendam\,\orcidlink{0000-0003-3953-9532}$^{1,\ddag}$}
\author{David~O.~Jones\,\orcidlink{0000-0002-6230-0151}$^{12}$}
\author{Michael~A.~Tucker\,\orcidlink{0000-0002-2471-8442}$^{4,5,\S}$}

\affiliation{$^{1}$ Institute for Astronomy, University of Hawai‘i, 2680 Woodlawn Drive, Honolulu, HI 96822, USA}

\affiliation{$^{2}$ Department of Astronomy, University of Illinois Urbana-Champaign, 1002 West Green Street, Urbana, IL 61801, USA}

\affiliation{$^{3}$ NSF-Simons AI Institute for the Sky (SkAI), 172 E. Chestnut St., Chicago, IL 60611, USA}

\affiliation{$^{4}$ Department of Astronomy, The Ohio State University, 140 West 18th Avenue, Columbus, OH 43210, USA}

\affiliation{$^{5}$ Center for Cosmology and AstroParticle Physics, The Ohio State University, 191 West Woodruff Avenue, Columbus, OH 43210, USA}
 
\affiliation{$^{6}$ Department of Physics, The Ohio State University, 191 West Woodruff Avenue, Columbus, OH 43210, USA}
 
\affiliation{$^{7}$ Department of Astronomy, School of Physics, Peking University, 5 Yiheyuan Road, Haidian District, Beijing 100871, People's Republic of China}
 
\affiliation{$^{8}$ Kavli Institute of Astronomy and Astrophysics, Peking University, 5 Yiheyuan Road, Haidian District, Beijing 100871, People's Republic of China}
 
\affiliation{$^{9}$ National Astronomical Observatories, Chinese Academy of Sciences, 20A Datun Road, Chaoyang District, Beijing 100101, China}
 
\affiliation{$^{10}$ Instituto de Estudios Astrof\'isicos, Facultad de Ingenier\'ia y Ciencias, Universidad Diego Portales, Av. Ej\'ercito Libertador 441, Santiago, Chile}
 
\affiliation{$^{11}$ School of Physics, The University of Melbourne, Parkville, VIC, Australia}

\affiliation{$^{12}$ Institute for Astronomy, University of Hawai‘i, 640 N. A‘ohoku Pl., Hilo, HI 96720, USA}
 
\email{Corresponding author: dddesai@hawaii.edu}
\altaffiltext{$\dagger$}{These authors contributed equally to this work.}
\altaffiltext{*}{NHFP Einstein Fellow}
\altaffiltext{$\ddag$}{NSF Graduate Research Fellow}
\altaffiltext{$\S$}{CCAPP Fellow}

\begin{abstract}
\noindent We present volumetric rates and luminosity functions (LFs) of tidal disruption events (TDEs) based on 21 TDEs from the All-Sky Automated Survey for Supernovae (ASAS-SN) between 2014 and 2024. We calculate completeness corrections from injection-recovery simulations of the ASAS-SN light curves using individual TDE templates. We find a total volumetric rate of $\log_{10}(R_\textrm{tot}\,[\Runit]) = 2.32^{+0.15}_{-0.23}$ ($R_\textrm{tot} = 210 \pm 85$~$\Runit$), for $z \leq 0.12$. We also measure the rates of two growing TDE subpopulations. The one off-nuclear TDE in our standard sample implies $\log_{10}(R_{\textrm{off-nuclear}}\,[\Runit]) = 0.39^{+0.52}_{-0.76}$ ($1.2^{+3.4}_{-1.0}$\% of the total rate). We measure the rate of recurrent TDEs (rTDEs) to be $\log_{10}(R_{\rm recurrent}\,[\Runit]) = 1.26^{+0.23}_{-0.62}$ ($9^{+8}_{-6}\%$ of the total rate) from the four rTDEs in our standard sample, increasing to $1.45^{+0.20}_{-0.35}$ ($13^{+10}_{-7}\%$) when the well-studied ASASSN-14ko is included. Galaxies with substantial ($>$0.1\%) burst-mass fractions and steeply falling star-formation histories host the majority of TDEs. We recover the known overrepresentation of TDEs in post-starburst galaxies, by a factor of $15^{+16}_{-11}$ relative to matched control galaxies, and a weaker enhancement of $5^{+5}_{-4}$ in quiescent Balmer-strong galaxies. TDE-H+He events dominate the volumetric TDE rate, with TDE-H, TDE-He, and TDE-Featureless events comprising $<$15\% of nearby TDEs. The $g$-band LFs are statistically indistinguishable when split at the median X-ray to UV/optical blackbody luminosity ratio of $\sim$1\%, corroborating unification models where the observed multi-wavelength emission is set primarily by viewing angle effects. Extrapolating the measured TDE LFs to higher luminosities and more massive black holes supports a TDE origin for luminous ambiguous nuclear transients and extreme nuclear transients. \\
\end{abstract}

\keywords{Astronomy data analysis (1858) --- Sky surveys (1464) --- Supermassive black holes (1663) --- Tidal disruption (1696) --- Time domain astronomy (2109)}
 
\maketitle

\section{Introduction} \label{sec:intro}

The vast majority of massive galaxies in the local universe host a supermassive black hole (SMBH) in their nucleus \citep[e.g.,][]{Kormendy1995, Magorrian1998, Ho2008, Gultekin2009, Kormendy2013}. A small fraction, $\sim$5\% \citep[e.g.,][]{Ho2008, Lacerda2020}, are accreting matter at a substantial rate and can be detected as active galactic nuclei \citep[AGNs; e.g.,][]{Rees1984, Antonucci1993, Reynolds1997, Stern2005, Kewley2006}. The remaining majority of inactive, or quiescent, SMBHs are nearly impossible to detect, save for a few dozen nearby and/or massive SMBHs for which stars \citep[e.g.,][]{Kormendy1996, Ghez2005, Gebhardt2011} and/or gas \citep[e.g.,][]{Ford1994, Atkinson2005} within the SMBH's sphere of influence can be resolved, and two where emission near the horizon has been imaged \citep{EventHorizonTelescope2019, EventHorizonTelescope2022}.
 
Fortunately, the recent growth of time-domain astronomy has provided a promising new probe of quiescent SMBHs, including many at cosmological distances. These are the flares resulting from a tidal disruption event (TDE), when a star passes too close to an SMBH and is ripped apart as its self-gravity is overwhelmed by the intense tidal forces of the SMBH \citep[e.g.,][]{Rees1988, Phinney1989, Evans1989, Ulmer1999}. TDE candidates have been observed across the electromagnetic spectrum \citep[e.g.,][]{Auchettl2017, Alexander2020, Gezari2021}, with many showing emission in multiple energy bands. Optical selection has proven particularly effective, with more than 100 optical TDEs discovered to date by surveys like the All-Sky Automated Survey for Supernovae \citep[ASAS-SN;][]{Shappee2014, Kochanek2017}, the Asteroid Terrestrial Impact Last Alert System \citep[ATLAS;][]{Tonry2018}, the Gravitational-wave Optical Transient Observer \citep[GOTO;][]{Dyer2024}, the Panoramic Survey Telescope and Rapid Response System \citep[Pan-STARRS;][]{Chambers2016}, and the Zwicky Transient Facility \citep[ZTF;][]{Bellm2019}.
 
Although TDEs provide undeniable evidence of a dormant SMBH, several complications hinder a robust connection between TDE observables and SMBH parameters. These include various properties of the disrupted star \citep[e.g.,][]{MacLeod2012b, Guillochon2013, Mockler2019, LawSmith2019} and the SMBH \citep[e.g.,][]{Ulmer1999, Mockler2019, Gafton2019} that affect the observed TDE emission. An accurate estimate of the rate at which TDEs occur is also fundamental to their use as probes of the SMBH population. Theoretically, TDE rates depend primarily on the diffusion of stars into the ``loss cone'' \citep[e.g.,][]{Frank1976, Lightman1977, Cohn1978, Stone2019}. Initially, there was tension between the theoretically predicted rate of TDEs \citep[$\sim$$10^{-4}$ yr$^{-1}$ galaxy$^{-1}$;][]{Magorrian1999, Wang2004, Stone2016} and the observed rate of TDEs in optical bands \citep[$\approx3 \times 10^{-5}$ yr$^{-1}$ galaxy$^{-1}$;][]{vanVelzen2014, Holoien2016a, Yao2023}. However, the increasing number of TDE candidates identified in other wavebands, such as the X-ray \citep[e.g.,][]{Sazonov2021, Grotova2025, EylesFerris2025}, infrared \citep[e.g.,][]{Masterson2024}, and radio \citep[e.g.,][]{Dykaar2024}, together with improved treatment of selection effects, helps alleviate this tension.
 
Detailed estimates of the volumetric TDE rate, particularly its dependence on SMBH mass, can shed light on important, broader questions about SMBHs. In the lower-mass intermediate-mass black hole (IMBH) regime, the SMBH mass function provides insight into BH seeding mechanisms in the early universe \citep[e.g.,][]{Ricarte2018, Greene2020}. At the high-mass end, since Solar-like stars are directly captured, TDE rate constraints can inform studies on the disruption of evolved \citep[e.g.,][]{MacLeod2012b} and massive stars \citep[e.g.,][]{Hinkle2025, Wiseman2025, Graham2026}, and the spin of high-mass quiescent SMBHs \citep[e.g.,][]{Kesden2012, Stone2019, Huang2024a}. Large TDE samples may enable population-level constraints on the spins of high-mass quiescent SMBHs \citep[e.g.,][]{Shvartzvald2024, Karmen2026}, which will complement AGN spin measurements \citep[e.g.,][]{Brenneman2006, Reynolds2021}. The TDE rate is relatively flat across SMBH masses in the range between $10^5$ M$_{\odot}$ and $\sim$$10^{7}$ M$_{\odot}$, falling sharply above $10^{7.5}$ M$_{\odot}$ \citep{Kochanek2016b, vanVelzen2018, Yao2023}. This decline is approximately consistent with the so-called ``Hills mass'' \citep{Hills1975} for Solar-like stars, where disruption occurs within the SMBH event horizon and is unobservable. This population-level trend suggests that the optical TDE population is dominated by the tidal disruption of low-mass ($\lesssim$2 M$_{\odot}$) main-sequence stars rather than non-TDE nuclear transients.

Existing estimates of the optical TDE luminosity function (LF) show a decline in the volumetric rate towards higher $g$-band luminosities. The decline is reasonably approximated by a power-law of $dN/dL_g \propto L^{-\alpha}$, with $\alpha$ between 2.3 and 3 \citep{vanVelzen2018, Lin2022, Charalampopoulos2023, Yao2023}. More detailed modeling by \citet{Yao2023} shows that the optical LF is better fit as a broken power-law with a break luminosity of $\sim$$10^{43}$ erg s$^{-1}$. The LF based on blackbody models for the bolometric luminosity also shows a decline at higher luminosities, with a slope of $\alpha \sim 2.4$ \citep{Lin2022, Yao2023}. The optical TDE LF is also broadly similar to the X-ray TDE LF \citep[e.g.,][]{Auchettl2018, Sazonov2021, Grotova2025}. TDE rates as a function of host-galaxy properties probe the processes producing TDEs. This includes understanding the overrepresentation of TDEs in quiescent Balmer-strong (QBS) and post-starburst (PSB) galaxies \citep[e.g.,][]{Arcavi2014, French2016, French2018, LawSmith2017, Graur2018a}, for which several models have been proposed \citep[e.g.,][]{Stone2018, Bortolas2022, Mockler2023, Melchor2024, Wang2024a, Teboul2025}, but no certain explanation exists \citep{Shepherd2026}.
 
ASAS-SN \citep{Shappee2014, Kochanek2017, Hart2023} provides an ideal dataset to measure the rate of transients. It has surveyed nearly the entire visible sky since 2014 in an untargeted fashion, reaching a roughly daily cadence following its expansion in late 2017. Additionally, the shallow depth ($m_g \lesssim 18$ mag) allows for high spectroscopic completeness \citep[][]{Holoien2017c, Holoien2017b, Holoien2017a, Holoien2019a, Neumann2023,Desai2026a}. This wide-area, long-baseline, and high-cadence coverage complements rate estimates from deeper surveys. ZTF \citep[e.g.,][]{Perley2020, Yao2023, Dimitriadis2025} observes only the northern sky, while ATLAS \citep[e.g.,][]{Srivastav2022, Srivastav2026} monitored only the northern sky until its southern units came online in 2022, and both have had longer average cadences than ASAS-SN since its expansion. Previous papers in this series have presented the rates of Type Ia supernovae and their subtypes \citep{Desai2024, Desai2026a}, core-collapse supernovae of various types (\citealt{Pessi2025}; Pessi et al., in prep.), and kilonovae \citep{Desai2026b}.
 
Here, we leverage the long history of successful ASAS-SN TDE discoveries to compute the local volumetric TDE rate. In Section \ref{sec:sample}, we describe the TDE sample, measurements of TDE and host-galaxy properties, and our refitting of the ASAS-SN light curves to compute peak apparent magnitudes. Section \ref{sec:rates} outlines the completeness corrections and rate computations. We next present the TDE rates and associated results on the $g$-band LF with respect to spectroscopic type and X-ray luminosity (Section \ref{sec:gband_LF}), their dependence on UV/optical blackbody parameters (Section \ref{sec:BB_rates}), and how they vary with the properties of their host galaxies (Section \ref{sec:hostgal_rates}). Finally, we summarize our conclusions in Section \ref{sec:summary}. Throughout this manuscript, we adopt a flat $\Lambda$CDM cosmology with $H_0 = 70~\mathrm{km~s^{-1}~Mpc^{-1}}$ and $\Omega_{m,0} = 0.3$, and define $h_{70} = H_0/(70~\mathrm{km~s^{-1}~Mpc^{-1}})$.

\section{The TDE Sample} \label{sec:sample}

In this section, we describe our sample of TDEs discovered or recovered by ASAS-SN (Section~\ref{subsec:asassn_catalog}) and the properties against which we compare their rates: the photometric and spectroscopic properties of the TDEs (Section~\ref{subsec:tde_properties}), the properties of their host galaxies (Section~\ref{subsec:host_properties}), and the masses of their central SMBHs (Section~\ref{subsec:smbh_masses}). We also detail the creation of our light curve templates, their application to ASAS-SN data (Section~\ref{subsec:LC_temp_fit}), and the cuts that define our standard sample for rate calculations (Section~\ref{subsec:std_sample}).

\subsection{ASAS-SN TDE Catalog} \label{subsec:asassn_catalog}

ASAS-SN has operated in two distinct phases: a $V$-band era (2012--late 2017) and a $g$-band era (late 2017--present). We begin our analysis in 2014, when ASAS-SN expanded to both hemispheres, and assign 2014--2017 to the $V$-band era and 2018--2024 to the $g$-band era. During the $V$-band era, ASAS-SN achieved a typical single-visit depth of $m_V \lesssim 17$~mag with a typical cadence of $\sim$2--3 days over much of the survey footprint. The transition to the $g$-band significantly improved survey performance, reaching a typical single-visit depth of $m_g \lesssim 18$~mag at a cadence of $\sim$1--2 days. The full ASAS-SN transient archive and its photometric pipeline are described in \citet{Hart2023}.

Our TDE sample spans the full analysis period from UTC 2014-01-01 to UTC 2024-12-31, covering both the $V$-band and $g$-band eras. We compiled TDEs that were either discovered by ASAS-SN or independently recovered in ASAS-SN data within this time range. To identify independently recovered events, we cross-matched all TDEs reported to the Transient Name Server\footnote{\url{https://www.wis-tns.org/}} (TNS) with the ASAS-SN transient archive\footnote{\url{https://www.astronomy.ohio-state.edu/asassn/transients.html}}. A TDE is considered ``recovered'' if another group first reported it, but it would have been independently flagged in the ASAS-SN data. In total, 11 TDEs were discovered by ASAS-SN and 14 were independently recovered, yielding 25 total TDEs. Tables~\ref{tab:tde_sample_tdeprops} and \ref{tab:tde_sample_hostprops} list the TDE and host-galaxy properties of the full sample of 25 TDEs, respectively, and Appendix~\ref{app:sources} gives the source of the data.

% TDE properties table
% Requires: \usepackage{booktabs}, \usepackage{graphicx}

\begin{table*}[t]
    \centering
    \footnotesize
    \renewcommand{\arraystretch}{1.3}
    \setlength{\tabcolsep}{4pt}
    \caption{Properties of the TDEs discovered or recovered in ASAS-SN (UTC 2014-01-01 to UTC 2024-12-31).}
    \label{tab:tde_sample_tdeprops}
    \resizebox{\textwidth}{!}{%
    \begin{tabular}{l l c c c c c c c c c c c c}
    \toprule
    {IAU Name} & {Discovery Name$^{a}$} & {RA} & {Dec} & {Spec.} & {$z$} & {$A_{g, \rm MW}$} & {$K(z)$} & {$m_{g,\text{peak}}$} & {$M_{g,\text{peak}}$} & {$\log_{10}\!\left(L_{\rm BB}/(\mathrm{erg\,s^{-1}})\right)$} & {$\log_{10}\!\left(T_{\rm BB}/\mathrm{K}\right)$} & {$\log_{10}\!\left(R_{\rm BB}/\mathrm{cm}\right)$} & {$\log_{10}\!\left(L_{\rm X}/(\mathrm{erg\,s^{-1}})\right)$} \\
     & & {[hh:mm:ss]} & {[dd:mm:ss]} & {Type} & & {[mag]} & {[mag]} & {[mag]} & {[mag]} &  &  &  &  \\
    \midrule
    \nodata & ASASSN-14ae$^{b}$ & 11:08:40.11 & +34:05:52.2 & TDE-H & 0.0436 & 0.058 & $-0.101$ & 16.96 & $-19.43$ & $43.96 \pm 0.02$ & $4.36 \pm 0.01$ & $14.83 \pm 0.02$ & $<41.1$ \\
    \nodata & ASASSN-14li$^{b}$ & 12:48:15.23 & +17:46:26.44 & TDE-H+He & 0.0206 & 0.085 & $-0.058$ & 16.43 & $-18.36$ & $44.08 \pm 0.11$ & $4.66 \pm 0.05$ & $14.29 \pm 0.11$ & $43.4$ \\
    \nodata & ASASSN-15oi$^{b}$ & 20:39:09.183 & $-$30:45:20.10 & TDE-He & 0.0479 & 0.225 & $-0.111$ & 16.18 & $-20.58$ & $44.29 \pm 0.02$ & $4.36 \pm 0.01$ & $15.00 \pm 0.02$ & $42.7$ \\
    2018dyb & ASASSN-18pg & 16:10:58.774 & $-$60:55:23.16 & TDE-H+He & 0.0179 & 0.763 & $-0.044$ & 15.42 & $-19.75$ & $44.237 \pm 0.004$ & $4.41 \pm 0.01$ & $14.88 \pm 0.01$ & $41.0$ \\
    2018fyk & ASASSN-18ul & 22:50:16.090 & $-$44:51:53.50 & TDE-H+He & 0.059 & 0.044 & $-0.159$ & 17.60 & $-19.39$ & $44.60 \pm 0.02$ & $4.59 \pm 0.01$ & $14.69 \pm 0.02$ & $43.6$ \\
    2018hyz & ASASSN-18zj & 10:06:50.871 & +01:41:34.08 & TDE-H & 0.0457 & 0.114 & $-0.101$ & 16.58 & $-19.97$ & $44.16 \pm 0.01$ & $4.32 \pm 0.01$ & $15.02 \pm 0.01$ & $41.4$ \\
    2018zr & PS18kh & 07:56:54.537 & +34:15:43.61 & TDE-H & 0.071 & 0.154 & $-0.107$ & 17.58 & $-19.99$ & $43.86 \pm 0.01$ & $4.13 \pm 0.01$ & $15.24 \pm 0.02$ & $41.6$ \\
    2019ahk & ASASSN-19bt & 07:00:11.546 & $-$66:02:24.14 & TDE-H & 0.0262 & 0.404 & $-0.052$ & 15.42 & $-20.22$ & $44.020 \pm 0.002$ & $4.230 \pm 0.001$ & $15.124 \pm 0.002$ & $40.8$ \\
    2019azh & ASASSN-19dj & 08:13:16.945 & +22:38:54.03 & TDE-H+He & 0.0223 & 0.146 & $-0.061$ & 15.50 & $-19.53$ & $44.380 \pm 0.004$ & $4.580 \pm 0.002$ & $14.604 \pm 0.004$ & $43.0$ \\
    2019dsg & ZTF19aapreis & 20:57:02.974 & +14:12:15.86 & TDE-H+He & 0.0512 & 0.340 & $-0.145$ & 17.93 & $-19.05$ & $44.44 \pm 0.02$ & $4.72 \pm 0.03$ & $14.36 \pm 0.06$ & $43.5$ \\
    2019qiz & ZTF19abzrhgq & 04:46:37.880 & $-$10:13:34.90 & TDE-H+He & 0.0151 & 0.372 & $-0.036$ & 15.89 & $-18.52$ & $43.53 \pm 0.01$ & $4.362 \pm 0.001$ & $14.615 \pm 0.004$ & $40.7$ \\
    2019vcb & ZTF19acspeuw & 12:38:56.380 & +33:09:57.28 & TDE-H+He & 0.088 & 0.058 & $-0.204$ & 18.46 & $-19.42$ & $44.13^{+0.12}_{-0.14}$ & $4.39^{+0.04}_{-0.05}$ & $14.84 \pm 0.08$ & $43.0$ \\
    2020neh & ZTF20abgwfek & 15:21:20.090 & +14:04:10.52 & TDE-H+He & 0.0620 & 0.146 & $-0.144$ & 17.92 & $-19.30$ & $43.78 \pm 0.01$ & $4.38 \pm 0.01$ & $14.71 \pm 0.01$ & $<41.7$ \\
    2020nov & ZTF20abisysx & 16:58:12.970 & +02:07:03.04 & TDE-H & 0.0826 & 0.373 & $-0.122$ & 18.12 & $-20.00$ & $43.37 \pm 0.10$ & $4.13 \pm 0.04$ & $15.00 \pm 0.05$ & $43.3$ \\
    2020vwl & Gaia20etp & 15:30:37.800 & +26:58:56.89 & TDE-H+He & 0.0325 & 0.158 & $-0.091$ & 17.59 & $-18.25$ & $43.90 \pm 0.16$ & $4.65 \pm 0.06$ & $14.23 \pm 0.15$ & $<41.3$ \\
    2022bdw & ATLAS22dth & 08:25:10.360 & +18:34:57.50 & TDE-H+He & 0.0378 & 0.155 & $-0.107$ & 17.51 & $-18.65$ & $44.30 \pm 0.10$ & $4.69 \pm 0.05$ & $14.35 \pm 0.05$ & $<41.6$ \\
    2022dbl & ASASSN-22ci & 12:20:45.010 & +49:33:04.68 & TDE-H+He & 0.0284 & 0.063 & $-0.075$ & 17.18 & $-18.28$ & $43.69 \pm 0.01$ & $4.51 \pm 0.03$ & $14.40 \pm 0.05$ & $41.5$ \\
    2022dsb & ASASSN-22cs & 15:42:21.743 & $-$22:40:14.04 & TDE-H+He & 0.0234 & 0.767 & $-0.068$ & 17.48 & $-18.27$ & $44.54 \pm 0.03$ & $4.75 \pm 0.03$ & $14.34 \pm 0.06$ & $43.3$ \\
    2022hvp & ZTF22aagyuao & 09:54:45.240 & +55:26:25.37 & TDE-He & 0.112 & 0.025 & $-0.284$ & 16.63 & $-21.69$ & $45.35 \pm 0.01$ & $4.535 \pm 0.002$ & $15.181 \pm 0.004$ & $<42.7$ \\
    2022lri & ATLAS22pnz & 02:20:08.009 & $-$22:43:15.42 & TDE-H+He & 0.0328 & 0.061 & $-0.094$ & 17.73 & $-18.02$ & $44.31 \pm 0.04$ & $4.72 \pm 0.05$ & $14.29 \pm 0.11$ & $44.2$ \\
    2023clx & ASASSN-23bd & 11:40:09.397 & +15:19:38.54 & TDE-H+He & 0.0111 & 0.103 & $-0.019$ & 16.46 & $-17.03$ & $42.740 \pm 0.004$ & $4.15 \pm 0.02$ & $14.65 \pm 0.03$ & $39.6$ \\
    2023mhs & ZTF23aaqdjhi & 13:43:15.681 & +19:15:00.90 & TDE-H+He & 0.0482 & 0.075 & $-0.089$ & 15.89 & $-20.74$ & $44.32 \pm 0.01$ & $4.202 \pm 0.001$ & $15.330 \pm 0.004$ & $42.4$ \\
    2024as & ATLAS24acr & 10:05:06.320 & +17:29:21.05 & TDE-H & 0.1225 & 0.120 & $-0.241$ & 18.53 & $-20.14$ & $43.85 \pm 0.01$ & $4.28 \pm 0.10$ & $14.95 \pm 0.19$ & $42.5$ \\
    2024pvu & ZTF18abxnvoz & 23:31:11.930 & +22:15:32.20 & TDE-H+He & 0.0488 & 0.198 & $-0.096$ & 17.18 & $-19.60$ & $43.951 \pm 0.002$ & $4.24 \pm 0.02$ & $15.07 \pm 0.04$ & $42.3$ \\
    2024tvd & ZTF22aaigqsr & 17:10:42.585 & +28:50:15.11 & TDE-H+He & 0.0449 & 0.164 & $-0.092$ & 17.13 & $-19.44$ & $43.719 \pm 0.003$ & $4.26 \pm 0.02$ & $14.92 \pm 0.05$ & $43.5$ \\
    \bottomrule
    \end{tabular}%
    }

    \vspace{4pt}
    \begin{minipage}{\textwidth}
    \raggedright
    Note. --- This table will be made available in machine-readable form on Zenodo upon publication. \\
    $^{a}$ Name assigned by the discovering survey. TDEs with an ``ASASSN-''name were discovered by ASAS-SN; the others were independently recovered in ASAS-SN data. \\
    $^b$ Discovered during the ASAS-SN $V$-band era (2014--2017). For these events, $m_{g,\text{peak}}$ is converted from the measured $V$-band peak magnitude using the $g-V$ color of the TDE's blackbody at peak (Section~\ref{subsec:LC_temp_fit}).\\
    Columns: $A_{g, \rm MW}$ is the $g$-band Milky Way line-of-sight extinction \citep{Schlafly2011}; $K(z)$ is the blackbody $K$-correction (Equation~\ref{eq:kcorr}); $m_{g,\text{peak}}$ is the peak apparent magnitude from fitting the ASAS-SN light curve with the per-event light curve templates (Section~\ref{subsec:LC_temp_fit}); $M_{g,\text{peak}}$ is the peak absolute magnitude from Equation~\ref{eq:absmag}.
    $L_{\rm BB}$, $T_{\rm BB}$, and $R_{\rm BB}$ are the UV/optical blackbody luminosity, temperature, and radius at peak (Section~\ref{subsec:tde_properties}).
    $L_{\rm X}$ is the maximum observed X-ray luminosity or 3$\sigma$ upper limit, as appropriate, for each TDE (Section~\ref{subsec:tde_properties}).
    \end{minipage}
\end{table*}
% Host-galaxy properties table
% Requires: \usepackage{booktabs}, \usepackage{graphicx}

\begin{table*}[t]
    \centering
    \footnotesize
    \renewcommand{\arraystretch}{1.5}
    \setlength{\tabcolsep}{4pt}
    \caption{Host-galaxy properties for the ASAS-SN TDE sample (UTC 2014-01-01 to UTC 2024-12-31).}
    \label{tab:tde_sample_hostprops}
    \resizebox{\textwidth}{!}{%
    \begin{tabular}{l l c c c c c c c c c c}
    \toprule
    {IAU Name} & {Discovery Name} & {$t_{\rm burst}$} & {$\log_{10}(M_{\star,\rm burst}/M_{\odot})$} & {$\log_{10}(M_{\star,\rm old}/M_{\odot})$} & {$\alpha$} & {$A_V$} & {$M_{\star,\rm burst}/M_{\star,\rm tot}$} & {$\log_{10}(M_{\rm BH}/M_{\odot})$} & {$n$} & {H$\alpha$ EW} & {Lick H$\delta_A$} \\
     & & {[Gyr]} & & & & {[mag]} & & & & {[\AA]} & {[\AA]} \\
    \midrule
    \nodata & ASASSN-14ae & $0.196^{+0.022}_{-0.014}$ & $7.89 \pm 0.03$ & $9.63 \pm 0.02$ & $763^{+134}_{-158}$ & $0.024^{+0.024}_{-0.014}$ & $(1.8 \pm 0.2)\times10^{-2}$ & $6.12$ & $1.09^{+0.02}_{-0.01}$ & $0.70 \pm 0.40$ & $3.40 \pm 0.80$ \\
    \nodata & ASASSN-14li & $1.235^{+0.007}_{-0.005}$ & $9.20^{+0.02}_{-0.01}$ & $3.78^{+1.29}_{-1.42}$ & $86 \pm 1$ & $0.012^{+0.012}_{-0.007}$ & $1.00 \pm {<}0.01$ & $6.14$ & $6.39^{+0.20}_{-0.22}$ & $0.60 \pm 0.50$ & $5.70 \pm 0.60$ \\
    \nodata & ASASSN-15oi & $0.104^{+0.010}_{-0.009}$ & $7.65 \pm 0.04$ & $10.15^{+0.03}_{-0.04}$ & $926^{+45}_{-58}$ & $0.314^{+0.080}_{-0.085}$ & $3.1^{+0.5}_{-0.4}\times10^{-3}$ & $5.86$ & $6.29^{+0.55}_{-0.78}$ & $-0.10 \pm 0.30$ & $1.90 \pm 0.70$ \\
    2018dyb & ASASSN-18pg & $1.716^{+0.804}_{-1.166}$ & $4.02^{+1.76}_{-1.96}$ & $10.58 \pm 0.05$ & $14^{+308}_{-14}$ & $1.975^{+0.019}_{-0.039}$ & $2.8^{+160}_{-2.7}\times10^{-7}$ & $6.78$ & \nodata$^{b}$ & $1.29 \pm 0.32$ & $-0.57 \pm 1.58$ \\
    2018fyk & ASASSN-18ul & $0.119 \pm 0.010$ & $8.14^{+0.04}_{-0.03}$ & $10.25^{+0.02}_{-0.01}$ & $741^{+78}_{-72}$ & $0.030^{+0.019}_{-0.015}$ & $7.6^{+0.7}_{-0.6}\times10^{-3}$ & $7.54$ & $4.13 \pm 0.02$ & $1.26 \pm 0.07$ & $-1.24 \pm 0.19$ \\
    2018hyz & ASASSN-18zj & $0.284^{+0.010}_{-0.012}$ & $8.47^{+0.01}_{-0.03}$ & $10.35^{+0.01}_{-0.02}$ & $817^{+50}_{-97}$ & $0.700^{+0.023}_{-0.044}$ & $(1.3 \pm 0.1)\times10^{-2}$ & $6.91$ & $4.55^{+0.78}_{-0.79}$ & $0.35 \pm 0.14$ & $5.43 \pm 0.41$ \\
    2018zr & PS18kh & $0.073^{+0.006}_{-0.004}$ & $7.07^{+0.07}_{-0.06}$ & $10.00^{+0.04}_{-0.03}$ & $961^{+27}_{-59}$ & $0.034^{+0.041}_{-0.022}$ & $(1.2 \pm 0.2)\times10^{-3}$ & $7.01$ & $1.18 \pm 0.01$ & $0.96 \pm 0.13$ & $-0.29 \pm 0.23$ \\
    2019ahk & ASASSN-19bt & $1.206^{+0.026}_{-0.025}$ & $8.57 \pm 0.02$ & $9.61 \pm 0.01$ & $674^{+144}_{-168}$ & $0.005 \pm 0.003$ & $(8.3 \pm 0.4)\times10^{-2}$ & $6.88$ & $7.95^{+0.03}_{-0.07}$ & $27.97 \pm 0.80$ & $6.08 \pm 0.92$ \\
    2019azh & ASASSN-19dj & $0.662^{+0.013}_{-0.010}$ & $9.25 \pm 0.02$ & $9.23^{+0.05}_{-0.07}$ & $135 \pm 2$ & $0.005^{+0.005}_{-0.003}$ & $0.51 \pm 0.04$ & $6.43$ & $5.41^{+0.19}_{-0.18}$ & $1.73 \pm 0.08$ & $7.67 \pm 0.17$ \\
    2019dsg & ZTF19aapreis & $1.884^{+0.514}_{-0.959}$ & $6.38^{+2.95}_{-3.51}$ & $10.36 \pm 0.09$ & $97^{+292}_{-94}$ & $0.365^{+0.078}_{-0.076}$ & $1.0^{+660}_{-1.0}\times10^{-4}$ & $6.88$ & $2.97^{+0.11}_{-0.10}$ & $11.53 \pm 1.42$ & $-0.82 \pm 0.11$ \\
    2019qiz & ZTF19abzrhgq & $0.500^{+0.019}_{-0.016}$ & $8.83^{+0.07}_{-0.08}$ & $7.83^{+0.26}_{-0.25}$ & $804^{+137}_{-186}$ & $1.993^{+0.005}_{-0.008}$ & $0.91^{+0.09}_{-0.15}$ & $5.42$ & $0.66 \pm 0.01$ & $0.02 \pm 0.92$ & $0.87 \pm 1.11$ \\
    2019vcb & ZTF19acspeuw & $1.701^{+0.870}_{-0.982}$ & $4.18^{+2.25}_{-2.15}$ & $10.20^{+0.02}_{-0.03}$ & $15^{+274}_{-14}$ & $1.704^{+0.050}_{-0.064}$ & $9.5^{+740}_{-9.4}\times10^{-7}$ & $6.03^c$ & $0.75^{+0.06}_{-0.04}$ & $15.13 \pm 0.86$ & $2.96 \pm 0.09$ \\
    2020neh & ZTF20abgwfek & $0.083^{+0.008}_{-0.006}$ & $8.36 \pm 0.03$ & $10.23 \pm 0.01$ & $875^{+59}_{-77}$ & $1.985^{+0.008}_{-0.010}$ & $(1.3 \pm 0.1)\times10^{-2}$ & $5.46$ & $1.56^{+0.17}_{-0.15}$ & $30.67 \pm 0.43$ & $1.86 \pm 0.94$ \\
    2020nov & ZTF20abisysx & $0.099^{+0.007}_{-0.006}$ & $8.58^{+0.02}_{-0.03}$ & $10.29^{+0.02}_{-0.01}$ & $897^{+55}_{-73}$ & $0.246^{+0.027}_{-0.036}$ & $(1.9 \pm 0.1)\times10^{-2}$ & $7.40^c$ & $4.62 \pm 0.06$ & $16.45 \pm 0.63$ & $3.58 \pm 0.63$ \\
    2020vwl & Gaia20etp & $0.143^{+0.033}_{-0.020}$ & $7.55^{+0.07}_{-0.06}$ & $9.89^{+0.03}_{-0.04}$ & $856^{+92}_{-117}$ & $0.373^{+0.070}_{-0.091}$ & $4.5^{+0.9}_{-0.7}\times10^{-3}$ & $6.08$ & $3.29^{+0.14}_{-0.13}$ & $0.04 \pm 0.20$ & $0.20 \pm 1.22$ \\
    2022bdw & ATLAS22dth & $0.110^{+0.006}_{-0.005}$ & $8.13 \pm 0.03$ & $9.92 \pm 0.02$ & $920^{+49}_{-47}$ & $0.284^{+0.075}_{-0.048}$ & $1.6^{+0.2}_{-0.1}\times10^{-2}$ & $6.19$ & $0.75^{+0.06}_{-0.04}$ & $9.41 \pm 0.33$ & $1.61 \pm 0.56$ \\
    2022dbl & ASASSN-22ci & $0.217^{+0.029}_{-0.020}$ & $7.85^{+0.06}_{-0.05}$ & $10.06^{+0.03}_{-0.02}$ & $744 \pm 132$ & $0.651^{+0.066}_{-0.057}$ & $6.1^{+0.9}_{-0.8}\times10^{-3}$ & $5.68$ & $1.52 \pm 0.03$ & $0.23 \pm 0.04$ & $2.15 \pm 0.53$ \\
    2022dsb & ASASSN-22cs & $0.371^{+0.010}_{-0.013}$ & $9.51^{+0.03}_{-0.04}$ & $10.21^{+0.03}_{-0.04}$ & $775^{+113}_{-111}$ & $0.421^{+0.076}_{-0.075}$ & $0.17 \pm 0.02$ & $6.58$ & $0.90 \pm 0.01$ & $13.87 \pm 0.76$ & $4.66 \pm 0.08$ \\
    2022hvp & ZTF22aagyuao & $1.512^{+0.960}_{-1.047}$ & $5.88^{+1.24}_{-3.08}$ & $10.72 \pm 0.06$ & $5^{+97}_{-5}$ & $0.068^{+0.086}_{-0.052}$ & $1.4^{+120}_{-1.4}\times10^{-5}$ & $7.47$ & $1.18^{+0.06}_{-0.05}$ & $0.51 \pm 0.24$ & $-1.02 \pm 0.93$ \\
    2022lri & ATLAS22pnz & $0.077 \pm 0.003$ & $7.19 \pm 0.02$ & $9.80 \pm 0.01$ & $971^{+18}_{-24}$ & $1.078 \pm 0.020$ & $(2.4 \pm 0.1)\times10^{-3}$ & $5.05^c$ & $1.59^{+0.04}_{-0.03}$ & $0.73 \pm 0.06$ & $-1.12 \pm 0.21$ \\
    2023clx & ASASSN-23bd & $0.131^{+0.016}_{-0.015}$ & $7.00^{+0.05}_{-0.07}$ & $9.28 \pm 0.05$ & $890^{+82}_{-122}$ & $0.013^{+0.017}_{-0.009}$ & $5.2^{+0.9}_{-0.8}\times10^{-3}$ & $4.63$ & $0.65 \pm 0.01$ & $3.37 \pm 0.35$ & $0.52 \pm 0.39$ \\
    2023mhs & ZTF23aaqdjhi & $0.097^{+0.004}_{-0.007}$ & $8.38^{+0.01}_{-0.02}$ & $10.40^{+0.02}_{-0.01}$ & $883^{+40}_{-39}$ & $1.953^{+0.033}_{-0.011}$ & $(9.5 \pm 0.4)\times10^{-3}$ & $6.19^c$ & $0.69^{+0.05}_{-0.02}$ & $13.92 \pm 0.32$ & $2.38 \pm 0.47$ \\
    2024as & ATLAS24acr & $0.174^{+0.018}_{-0.021}$ & $8.98^{+0.03}_{-0.04}$ & $10.49 \pm 0.03$ & $765^{+102}_{-119}$ & $0.357^{+0.086}_{-0.053}$ & $(3.0 \pm 0.3)\times10^{-2}$ & $6.96^c$ & $2.87^{+0.46}_{-0.41}$ & $22.08 \pm 0.95$ & $6.20 \pm 0.42$ \\
    2024pvu & ZTF18abxnvoz & $1.453^{+0.038}_{-0.039}$ & $10.19 \pm 0.06$ & $4.68^{+2.62}_{-2.29}$ & $70 \pm 3$ & $1.015^{+0.096}_{-0.105}$ & $1.00 \pm {<}0.01$ & $7.90^c$ & $2.82 \pm 0.03$ & $0.63 \pm 0.19$ & $4.03 \pm 0.39$ \\
    2024tvd$^a$ & ZTF22aaigqsr & \nodata & \nodata & \nodata & \nodata & \nodata & \nodata & $5.50$ & \nodata & \nodata & \nodata \\
    \bottomrule
    \end{tabular}%
    }

    \vspace{4pt}
    \begin{minipage}{\textwidth}
    \raggedright
    Note. --- This table will be made available in machine-readable form on Zenodo upon publication. \\
    $^a$ Host-galaxy properties measured at the nucleus are not applicable for the off-nuclear TDE~2024tvd. \\
    $^b$ No reliable S\'{e}rsic fit was possible for the host of TDE~2018dyb, which lies at low Galactic latitude in a crowded field; this event is also excluded from the standard sample by the $|b| > 15\degr$ cut (Section~\ref{subsec:std_sample}). \\
    $^c$ SMBH mass estimated from host-galaxy scaling relations. \\
    Columns: $t_{\rm burst}$, $M_{\star,\rm burst}$, $M_{\star,\rm old}$, $\alpha$, and $A_V$ are the \texttt{Bagpipes} burst age, stellar mass formed in the recent burst, stellar mass formed in the old component, declining slope of the burst SFH, and dust attenuation (Section~\ref{subsec:host_properties}).
    $M_{\rm BH}$ is the SMBH mass (Section~\ref{subsec:smbh_masses}); $n$ is the bulge S\'ersic index; H$\alpha$ EW and Lick H$\delta_A$ are the H$\alpha$ emission equivalent width and Lick H$\delta_A$ absorption index.
    \end{minipage}
\end{table*}

We individually inspected every TDE in our sample and confirmed both its TDE nature and its spectroscopic type. Based on this, we remove TDE 2020afhd (ZTF20abwtifz), which has also been reclassified as a Bowen Fluorescence Flare \citep{Arcavi2024}. Bowen Fluorescence Flares \citep[e.g.,][]{Trakhtenbrot2019a, Gromadzki2019, Makrygianni2023} are characterized by strong \ion{He}{2}\,$\lambda$4686 emission and the fluorescent Bowen lines \citep[e.g., \ion{O}{3}\,$\lambda$3133, \ion{O}{3}\,$\lambda$3760, \ion{N}{3}\,$\lambda$4100, and \ion{N}{3}\,$\lambda$4640, with most lines being blends;][]{Bowen1934, Bowen1935, Netzer1985}. They have primarily been observed in pre-existing AGNs and have been proposed to occur from enhanced accretion onto the SMBH, possibly from a TDE or unrelated perturbation to the accretion flow \citep[e.g.,][]{Trakhtenbrot2019a, Makrygianni2023}. Bowen Fluorescence Flares have an uncertain physical origin \citep[e.g.,][]{Sniegowska2025, Veres2026b} and share similarities with ambiguous nuclear transients \citep[ANTs;][]{Neustadt2020, Hinkle2022, Hinkle2024b, Wiseman2025}, which may also be TDEs occurring in AGN hosts. Although TDEs may power a subset of these exotic accretion flares, we restrict our estimate of the volumetric rates to the well-established class of UV/optical TDEs. This exclusion also applies to events like ASASSN-14ko, which is likely a repeating partial TDE \citep[e.g.,][]{Payne2021, Payne2022, Payne2023}, and ASASSN-15lh, which may be a superluminous supernova \citep[e.g.,][]{Dong2016, Godoy-Rivera2017}. Neither event is a standard UV/optical TDE based on their observational signatures.

\subsection{TDE Properties} \label{subsec:tde_properties}
 
We adopted spectroscopic classifications from TNS and the literature, primarily \citet{vanVelzen2020}. Optical TDE spectra are commonly grouped into four spectroscopic types \citep[e.g.,][]{Leloudas2019, vanVelzen2021, Hammerstein2023} defined by the broad emission features present in their optical spectra. TDE-H events are dominated by broad hydrogen Balmer emission (e.g., H$\alpha$, H$\beta$), while TDE-He events are dominated by broad \ion{He}{2}\,$\lambda$4686 emission with no clearly detected broad Balmer component. TDE-H+He events exhibit both broad hydrogen and helium features, and may also show Bowen fluorescence features, including blended \ion{N}{3} and \ion{O}{3}, while TDE-Featureless events exhibit a smooth, blue continuum with no discernible broad transient features \citep[e.g.,][]{Leloudas2019, vanVelzen2021, Hammerstein2023, Yao2026b}. Some TDEs show significant temporal evolution in their spectra \citep[e.g.,][]{Nicholl2019, Short2020, Hinkle2021a}; in these cases, we adopt the spectroscopic classification from the spectrum obtained closest to the optical peak. Our sample contains 6~TDE-H, 2~TDE-He, 17~TDE-H+He events, and no TDE-Featureless events.

We fit the TDE UV/optical SEDs from the Swift Ultraviolet and Optical Telescope \citep[UVOT;][]{Roming2005}, following standard procedures \citep[e.g.,][]{Hinkle2021b, Hinkle2022, Holoien2022}. In brief, we corrected the photometry for Galactic extinction, subtracted the host-galaxy flux in each band, and modeled the SED as a single-temperature blackbody (BB), sampling using a Markov Chain Monte Carlo (MCMC) method implemented in \texttt{emcee} \citep{ForemanMackey2013} with broad, flat priors on temperature ($1000 \textrm{ K} < T_{\textrm{BB}} < 60,000 \textrm{ K}$) and radius ($10^{13} \textrm{ cm} < R_{\textrm{BB}} < 10^{17} \textrm{ cm}$). We created a bolometric light curve from a well-sampled ground-based light curve by linearly interpolating a bolometric correction between Swift epochs and assuming a constant temperature outside of the Swift coverage. We propagated the uncertainty on the assumed constant temperature into the bolometric light curve.

We estimated the peak UV/optical blackbody luminosity by fitting a Gaussian Process (GP) to each bolometric light curve. We adopted the maximum of the GP posterior mean as the peak luminosity, with its uncertainty from the GP posterior at that time, and estimated the uncertainty on the time of peak by bootstrap resampling of the light curve. When at least one blackbody-fit epoch fell on each side of the time of peak luminosity, we similarly fit a GP to the temperature and radius time series and evaluated them at peak. For several TDEs, Swift observations are only available after peak, notably $\sim$144 days post-peak for TDE~2022lri. In these cases, because most TDEs exhibit comparatively modest temperature evolution \citep[e.g.,][]{vanVelzen2021, Hinkle2020a, Stein2026}, we instead adopted the temperature of the earliest available Swift epoch as the temperature at peak. Assuming a constant temperature between peak and the first Swift epoch, we derived the corresponding blackbody radius at peak luminosity using the Stefan-Boltzmann law.
 
In addition to the UV/optical parameters, we computed rest-frame unabsorbed 0.3--10 keV X-ray luminosities or 3$\sigma$ upper limits for each TDE. When possible, we collected the maximum X-ray luminosity from the literature. Because the X-ray coverage of TDEs is heterogeneous, the maximum X-ray luminosity may not represent the intrinsic peak X-ray luminosity for every event. For the few sources without published X-ray constraints, we used the online Swift XRT tools\footnote{\url{https://www.swift.ac.uk/user_objects/}} to measure an X-ray count rate or upper limit from the stacked coverage. Adopting the Galactic $n_H$ \citep{HI4PI2016}, we converted the estimated count rate to a flux assuming an absorbed 50 eV blackbody model, a commonly adopted approximation for soft X-ray TDE emission \citep[e.g.,][]{Auchettl2017, Kara2018, Saxton2020}, using \texttt{WebPIMMS} \citep{Mukai1993}.

\subsection{Host-Galaxy Properties} \label{subsec:host_properties}

We fit optical spectra of the TDE host galaxies with \texttt{Bagpipes} to estimate the properties of any recent burst of star formation. \texttt{Bagpipes} \citep{Carnall2018, Carnall2019} uses the \texttt{MultiNest} nested sampling algorithm \citep{Feroz2019} and the \citet{Bruzual2003} stellar population models. We extrapolated metallicity priors from the galaxy's stellar mass by using the stellar mass--metallicity trend in Table 2 of \citet[][]{Gallazzi2005}. Full details on the \texttt{Bagpipes} priors and the complete fit instructions are in \citet{Shepherd2026}.
 
We adopted a two-component star formation history (SFH): an old stellar component modeled by a delayed exponential ($\textrm{SFR}(t) \propto t \times e^{-t/\tau}$; \citealt{Simha2014}) and a young stellar component modeled by a double power-law ($\textrm{SFR}(t) \propto [(t/\tau)^{\alpha} + (t/\tau)^{-\beta}]^{-1}$; \citealt{Wild2020}) representing either a burst or an ongoing period of star formation. Here, $t$ is the age of the galaxy, $\tau$ is the e-folding time, $\alpha$ is the falling slope, and $\beta$ is the rising slope. \citet{Shepherd2026} contains more details about the choice of the SFH functional form and other \texttt{Bagpipes} parameters.
 
The equivalent width (EW) of the H$\alpha$ emission line and the Lick H$\delta_{\rm A}$ index from the host-galaxy spectra also hold important information on recent star formation. When possible, we adopt literature values \citep[e.g.,][]{French2016, Graur2018a, French2020a, Shepherd2026} for the TDE host galaxies. Otherwise, we calculated the H$\alpha$ emission line EW using \texttt{Bagpipes} (see \citealt{Shepherd2026}) and the Lick H$\delta_{\rm A}$ absorption line index using \texttt{pyLick} \citep{Borghi2022}. To facilitate a robust comparison, we created a control sample of galaxies from SDSS, whose H$\alpha$ EWs and H$\delta_{\rm A}$ indices were included in SDSS {\tt galSpec} \citep{Kauffmann2003a, Brinchmann2004, Tremonti2004}. The control sample was constructed by splitting the redshift range of the TDE hosts into five equal bins and splitting the stellar mass range of the TDE hosts into seven equal bins. We randomly selected 950 non-TDE-host galaxies per TDE host galaxy falling in each cell, such that a cell containing $n$ TDE hosts receives $950 \times n$ control galaxies. This yields $19{,}000$ control galaxies across the cells.
 
In addition to modeling recent star formation in the TDE hosts, we estimated morphological parameters that may correlate with the physical processes that produce TDEs. We measured the host-galaxy bulge S\'ersic indices using $r$-band imaging from the DESI Legacy Imaging Surveys \citep{Dey2019}. We were primarily interested in the bulge light profile, since the disrupted stars come from near the SMBH. We use the \texttt{sersic\_exp} profile model of \texttt{PyS\'ersic} \citep{Pasha2023}, which decomposes the galaxy into an inner S\'ersic component and an outer exponential component with the S\'ersic index fixed to $n=1$, representing the bulge and disk, respectively. The remaining analyses use the bulge S\'ersic index. One caveat of the \texttt{sersic\_exp} profile model is that it cannot capture higher-order components such as bars, arms, or asymmetrical structures.
 
For each \texttt{PyS\'ersic} fit, we provided the $r$-band galaxy image, its uncertainties, the PSF model, and a source mask. We used the $r$-band inverse-variance maps from the Legacy Survey to construct the uncertainties. Bad pixels in the inverse-variance maps were assigned the median value of the remaining pixels. We used \texttt{Photutils} \citep{Bradley2025} segmentation and deblending to identify the host galaxy and mask contaminants. The model was convolved with the local $r$-band Legacy Survey PSF in \texttt{PyS\'ersic} before comparison with the image. We estimated the posterior distributions of the model parameters using the Laplace approximation implemented in \texttt{PyS\'ersic} and report the median of the posterior distribution as the adopted bulge S\'ersic index.

\subsection{Supermassive Black Hole Masses} \label{subsec:smbh_masses}

When available, we adopt SMBH mass estimates inferred from late-time TDE emission \citep{Mummery2024a, Mummery2025}. These estimates use TDE observables directly, although they remain model-dependent and may have systematic uncertainties distinct from those of host-galaxy scaling relations. SMBH masses inferred from late-time plateau emission are consistent with expectations from host-galaxy stellar mass \citep[e.g.,][]{Mummery2024a, Mummery2025, Guolo2025}, although studies have shown that SMBH masses estimated from TDE emission may be underestimated \citep{Angus2026}.

Most of our SMBH mass estimates come from \citet{Mummery2024a} or \citet{Mummery2025} using the late-time plateau emission. For the few sources without existing estimates, typically recent events without extensive late-time monitoring, we use host-galaxy estimates from the M-$\sigma$ relation \citep[e.g.,][]{Gultekin2009, Kormendy2013} or stellar-mass scalings \citep[e.g.,][]{McConnell2013, Reines2015}. For all methods, the scatter on the SMBH mass estimates is at least 0.3 dex, and typically higher for the host-galaxy scalings. Nevertheless, the large mass bins used for the analysis and the significant statistical uncertainties make this a subdominant source of uncertainty in our TDE SMBH mass function.

\subsection{Light Curve Templates and Fitting} \label{subsec:LC_temp_fit}

TDE light curves exhibit diversity that no universal template can cleanly reproduce \citep[e.g.,][]{vanVelzen2021, Hammerstein2023}. Therefore, rather than adopting a single class-wide template as in the supernova analyses of Papers~I--III \citep{Desai2024, Pessi2025, Desai2026a}, we construct an individual template for each of the 25 TDEs in our sample. This per-TDE approach serves two purposes: it can provide more robust peak magnitude estimates than local interpolation when the peak is not well sampled, and it allows the injection-recovery simulations described in Section~\ref{subsec:phot_compl} to compute the completeness individually for each event, capturing the full diversity of light curve morphologies. The sample is also small enough that running separate completeness simulations for each TDE is computationally tractable.

Each template was constructed by fitting the available $g$-band or $V$-band photometry. These data come primarily from ASAS-SN or the Zwicky Transient Facility \citep[ZTF;][]{Bellm2019}, but in some cases, we used published photometry from individual discovery papers \citep[e.g.,][]{Holoien2019b, Holoien2020}. We fit the data with the Modular Open Source Fitter for Transients \citep[MOSFiT;][]{Guillochon2017b, Mockler2019} TDE model. An advantage of this approach is that we can extrapolate for events with either no pre-peak data and/or poor temporal coverage to both early and late times using the physically motivated, yet model-dependent, light-curve shape of the MOSFiT TDE model. The resulting normalized templates capture the full rise and decline morphology of each event and are scaled in the injection-recovery simulations by varying only the peak flux and time of peak.
 
\begin{figure}
    \centering
    \includegraphics[width=\columnwidth]{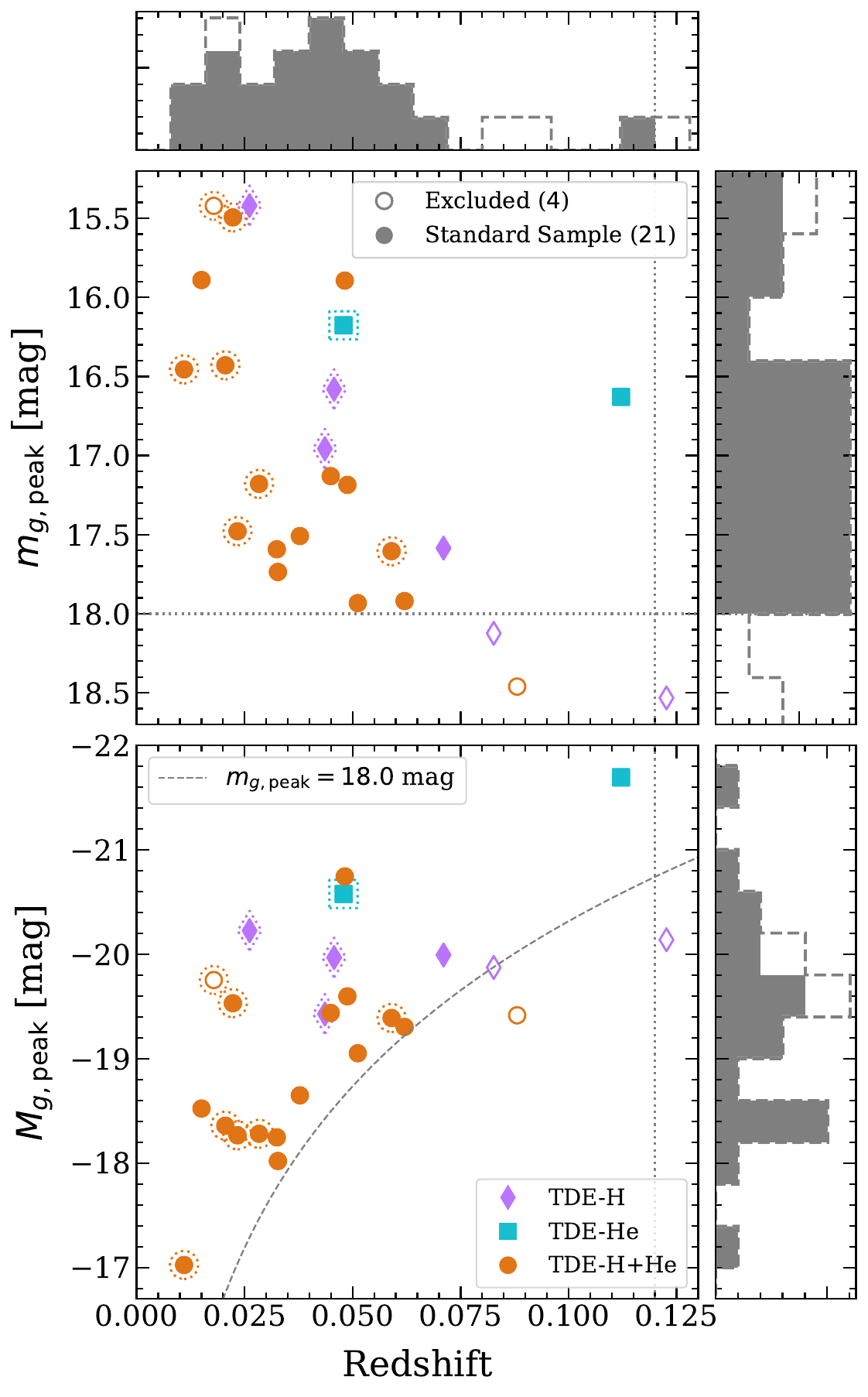}
    \caption{Peak apparent (top) and absolute (bottom) magnitude versus redshift distributions of the 25 TDEs discovered or recovered by ASAS-SN. The filled markers and filled histograms show our standard sample of 21 TDEs, while open markers and dashed histograms include the TDEs excluded by our standard cuts on Galactic latitude $b$, peak apparent magnitude $m_{g,\mathrm{peak}}$, and redshift $z$, as marked by the gray dotted lines. Marker shapes and color give the spectral classifications: purple diamonds for TDE-H, cyan squares for TDE-He, and orange circles for TDE-H+He. Points with an outer dotted shape denote ASAS-SN-discovered TDEs. The gray dashed curve in the bottom panel marks the envelope corresponding to $m_{g,\mathrm{peak}}=18.0$~mag based on the distance modulus for the redshift.}
    \label{fig:scatter_hist}
\end{figure}

To measure the ASAS-SN peak magnitude for each TDE, we fit its ASAS-SN light curve with the template derived from that same event using $\chi^2$ minimization with two free parameters: the time of peak, $t_\mathrm{peak}$, and the flux at peak, $F_\mathrm{peak}$. The resulting peak flux is converted to a peak apparent magnitude, $m_\mathrm{peak}$.

We compute absolute magnitudes for all TDEs using
\begin{equation}
    M_{\text{peak}} = m_{\text{peak}} - \mu - A_{\text{MW}} - K(z),
    \label{eq:absmag}
\end{equation}
where $m_{\text{peak}}$ is the apparent magnitude from light-curve fits in the respective discovery band, $\mu$ is the distance modulus computed from the redshift using the \texttt{astropy.cosmology} package \citep{Astropy2013, Astropy2018, Astropy2022}, and $A_{\text{MW}}$ is the Milky Way foreground extinction from \citet{Schlafly2011}, assuming a \citet{Fitzpatrick1999} extinction law with $R_V = 3.1$. We do not correct for host-galaxy extinction, but we examine the dependence of the rate on host-galaxy dust in Section~\ref{subsec:burst_props}. We compute the $K$-correction for each TDE from its peak blackbody temperature $T_{\rm BB}$ (Section~\ref{subsec:tde_properties}), evaluated at the $g$-band effective frequency $\nu_g$ ($\lambda_{g,\mathrm{eff}} \approx 4770$\,\AA):
\begin{equation}
    K(z) = -2.5\log_{10}\left[(1+z)\,\frac{B_\nu\left((1+z)\,\nu_g,\, T_{\rm BB}\right)}{B_\nu\left(\nu_g,\, T_{\rm BB}\right)}\right],
    \label{eq:kcorr}
\end{equation}
where $B_\nu$ is the Planck function \citep[e.g.,][]{Hogg2002}. For the $V$-band discovered TDEs, we likewise convert to $g$-band magnitudes using the $g-V$ color of the redshifted blackbody, with $\lambda_{V,\mathrm{eff}} \approx 5448$\,\AA.

\subsection{Standard Sample} \label{subsec:std_sample}

First, we require $m_{g,\mathrm{peak}} < 18.0$~mag (for $V$-band-era events, converted to $g$ as in Section~\ref{subsec:LC_temp_fit}), which corresponds to the regime where our photometric completeness is well-characterized by the injection-recovery simulations (Section~\ref{subsec:phot_compl}). This removes TDE 2019vcb (ZTF19acspeuw), TDE 2020nov (ZTF20abisysx), and TDE 2024as (ATLAS24acr). Second, we require a Galactic latitude $|b| > 15\degr$ to avoid regions of high and spatially variable Milky Way extinction where the photometric selection function is poorly constrained, following the Galactic-latitude criterion adopted in previous papers in this series \citep{Desai2024,  Desai2026a, Desai2026b, Pessi2025}. This removes TDE 2018dyb (ASASSN-18pg). Third, we restrict the redshift range to $0.0 < z \leq 0.12$. No TDEs in our sample lie at $z < 0.011$, so peculiar-velocity errors in the distance estimates are at most $\sim$8\% \citep[e.g.,][]{Tully2016}, much smaller than the $\gtrsim$40\% statistical uncertainties on our rates. We chose the upper limit of $z = 0.12$ which includes every TDE passing the other two cuts. It defines the outer boundary of our effective survey volume, to which the volume correction of each TDE is normalized (Section~\ref{subsec:phot_compl}). After applying these cuts, our standard sample consists of 21 TDEs.  
 
Figure~\ref{fig:scatter_hist} shows the peak apparent and absolute magnitude distributions as a function of redshift for the full, classification-vetted 25 TDE sample. The standard sample of 21 TDEs is shown with filled markers and excluded events are shown with open markers.

\section{Rate Computations} \label{sec:rates}

To compute volumetric TDE rates and LFs, we must correct the observed sample for the survey's incomplete detection and recovery of events. We follow the injection-recovery framework established in Papers~I--III \citep{Desai2024, Pessi2025, Desai2026a} and apply three completeness corrections: a photometric correction $F_1$, the probability that ASAS-SN recovers a TDE at its observed peak magnitude; a volume correction $F_2$, the fraction of the volume out to $z_\mathrm{max} = 0.12$ within which it could have been selected; and a spectroscopic correction $F_3$, the fraction of nuclear transients that receive a spectroscopic classification. Each lies between 0 and 1, and each TDE is weighted by $(F_1 F_2 F_3)^{-1}$. Section~\ref{subsec:phot_compl} describes $F_1$ and $F_2$, and Section~\ref{subsec:spec_compl} describes $F_3$. Section~\ref{subsec:vol_rates} combines them to derive the volumetric rates and Section~\ref{subsec:offnuc_rec} gives the rates of off-nuclear and recurrent TDEs.

\subsection{Photometric Completeness} \label{subsec:phot_compl}
 
\begin{figure}
    \centering
    \includegraphics[width=\columnwidth]{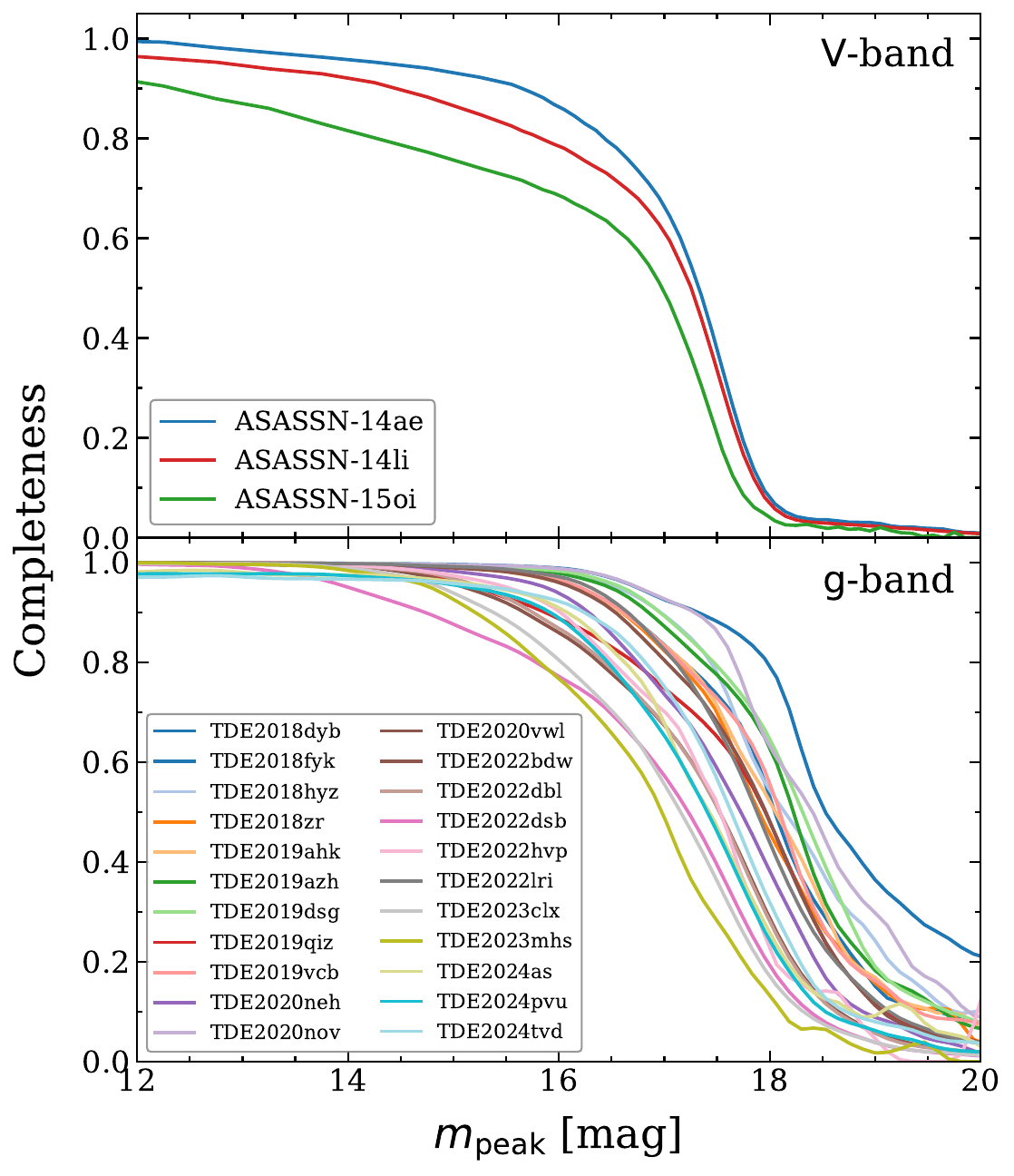}
    \caption{Photometric completeness ($F_1$) as a function of peak apparent magnitude for each TDE in our sample, shown separately for the $V$-band era (top) and the $g$-band era (bottom). Each curve corresponds to an individual TDE and uses the per-TDE light curve template described in Section~\ref{subsec:LC_temp_fit}. }
    \label{fig:completeness}
\end{figure}

To estimate the photometric completeness $F_1$ for each TDE, we employ an injection-recovery simulation following the methodology of Papers~I--III. We extract ASAS-SN light curves at $N_\mathrm{LC} = 100{,}000$ random sky coordinates distributed uniformly across the sky, and compute $F_1$ using only those with $|b| > 15\degr$, matching the Galactic-latitude cut of our standard sample (Section~\ref{subsec:std_sample}). These baseline light curves empirically capture the true observing conditions of the survey -- including photometric noise, observing cadence, weather interruptions, and seasonal gaps -- without tracing any specific astrophysical source. Using random sky positions rather than galaxy centers has a negligible effect on the completeness. Bright transients have signal-to-noise ratios high enough that host flux does not change whether an epoch is detected, while for faint transients the noise is dominated by the sky rather than the host. For each sky position, we inject the per-TDE light curve templates described in Section~\ref{subsec:LC_temp_fit} across $N_{\rm trial} = 100$ trials. For each trial, the redshift is drawn uniformly from the comoving volume out to the redshift at which the template would reach the simulation's limiting magnitude, so that the injected events span the range of peak apparent magnitudes of interest, and the peak time $t_\mathrm{peak}$ is drawn uniformly from the survey baseline. The injected light curves account for cosmological time dilation at each trial redshift. A trial is considered recovered if at least one epoch satisfies the discovery or recovery criterion described below.

The single-epoch recovery probability $p(\mathrm{SNR})$, the probability that an epoch with a given SNR yields a detection that passes the ASAS-SN pipeline and vetting, differs between the two survey eras. For the $V$-band era (2014--2017), we adopt the piecewise function derived in Paper~I \citep{Desai2024}, which saturates at $p = 0.65$ for $\mathrm{SNR} \geq 12$. For the $g$-band era (2018--2024), we use the per-year skew-normal efficiency models from Paper~III \citep{Desai2026a}, which peak at $p \approx 0.48$ averaged across all years and account for year-to-year changes in survey operations and transient-selection procedures, particularly after the onset of the COVID-19 pandemic. By modeling $p$ as a continuous function of SNR, this approach captures pipeline systematics and vetting efficiency that a simple flux threshold would miss.
 
Each TDE is assigned its own unique template rather than a class-wide template. This is motivated by the substantial morphological diversity of TDE light curves (Section~\ref{subsec:LC_temp_fit}), and it means that $F_1$ is evaluated individually per event. TDE light curves are also considerably longer-lived than SNe or kilonovae, with rise-and-fall timescales of weeks to months. This extended duration generally increases the number of epochs above threshold for a given peak flux, improving completeness relative to shorter-lived transients at comparable peak magnitudes.

For each injection, we record its peak apparent magnitude and whether it was recovered. $F_1(m_\mathrm{peak})$ is then the fraction of injections recovered in bins of peak apparent magnitude, i.e., the probability that ASAS-SN would have recovered the TDE at its observed brightness, and we evaluate it at the measured $m_\mathrm{peak}$ of each TDE. Figure~\ref{fig:completeness} shows the resulting $F_1$ completeness curves for all TDEs in our sample, separated by survey era. The spread among curves at a given magnitude reflects the diversity of TDE light curve morphologies, with broader and more slowly evolving events achieving higher completeness at fainter apparent magnitudes. 

$F_1$ corrects for recovery at a given apparent brightness, but not for the volume within which a TDE of that luminosity could have been selected. A TDE that falls below our magnitude limit beyond $z_{\mathrm{lim},i} < z_\mathrm{max}$ samples only part of the survey volume, so, following the $1/V_\mathrm{max}$ approach of Papers~I--III, we compute for each TDE
\begin{equation}
    F_{2,i} = \min\left(1,\, \frac{V\left(z_{\mathrm{lim},i}\right)}{V\left(z_{\mathrm{max}}\right)}\right),
    \label{eq:F2}
\end{equation}
where $z_{\mathrm{lim},i}$ is the redshift at which the $i^{\mathrm{th}}$ TDE, with $M_{\mathrm{peak},i} + A_{\mathrm{MW},i} + K(z_i)$, would have $m_\mathrm{peak} = m_\mathrm{lim}$, with $m_\mathrm{lim} = 18.0$~mag and $z_\mathrm{max} = 0.12$. This normalizes the effective detection volume of every TDE to the common maximum redshift of our sample.

\subsection{Spectroscopic Completeness} \label{subsec:spec_compl}

Not all transients discovered or recovered by ASAS-SN receive spectroscopic classification. To account for this incompleteness, we compute a spectroscopic completeness correction $F_3$ using the population of nuclear transients in the ASAS-SN archive as a proxy for the spectroscopic completeness of TDEs specifically.

We construct a sample of nuclear transients from two sources. First, from the full sample of 3355 ASAS-SN-discovered or recovered $g$-band transients in Paper~III \citep{Desai2026a}---which includes both classified and unclassified non-stellar objects with a host galaxy association, but excludes spectroscopically classified TDEs---we select the 771 events that lie within $2\arcsec$ of a host galaxy nucleus, as determined by cross-matching with the REGALADE galaxy catalog \citep{Tranin2026}. Second, we add the 25 ASAS-SN-discovered or recovered TDEs from our sample, none of which are in the Paper~III sample, yielding a combined sample of 796 nuclear transients. Of these, 662 ($83\%$) are spectroscopically classified.

We define $F_3$ for a given year as the ratio of spectroscopically classified to total nuclear transients in that year. For the $V$-band era (2014--2017), the ASAS-SN sample was nearly spectroscopically complete \citep{Holoien2017a, Holoien2017b, Holoien2017c, Holoien2019a}, with $F_3 = 1.00$, $1.00$, $1.00$, and $0.97$ for 2014 through 2017, respectively. For the $g$-band era, $F_3 = 0.86$, $0.87$, $0.73$, $0.85$, $0.69$, $0.83$, and $0.77$ for 2018 through 2024, respectively. The decline in later years mirrors the pattern seen in Paper~III, driven by the fainter $g$-band limiting magnitude, disruption of follow-up resources during the COVID-19 pandemic, and the reduced availability of spectroscopic follow-up for southern targets. Since the statistical uncertainties from the small TDE sample size dominate the total error budget, precise characterization of $F_3$ is not critical, and this approach provides a sufficiently accurate estimate.

\subsection{Volumetric Rates} \label{subsec:vol_rates}
 
The final statistical weight for the $i^{\mathrm{th}}$ TDE is $w_i = \left(F_{1,i}\, F_{2,i}\, F_{3,i}\right)^{-1}$. $F_1$ depends on the discovery filter because the single-epoch detection efficiency differs between the $V$-band and $g$-band eras. $F_2$ and $F_3$ are filter-independent. The combined volumetric rate is then
\begin{equation}
    R_{\mathrm{tot}} = \frac{\sum_{j=1}^{N_V} w_{j,V} + \sum_{i=1}^{N_g} w_{i,g}}{V \Delta t_{\mathrm{tot}} (1 - \sin b_{\lim})} ,
    \label{eq:rate_combined}
\end{equation}
where $V = \frac{4}{3}\pi d_\mathrm{max}^3$ is the total comoving volume defined by our redshift limit ($z_{\rm max} = 0.12$), $\Delta t_\mathrm{tot} = 11.0$~yr is the full survey baseline from UTC 2014-01-01 to UTC 2024-12-31, and $(1 - \sin b_{\lim})$ is the sky fraction correction for our Galactic latitude cut of $|b| > 15\degr$.

We estimate statistical uncertainties on all rates via bootstrapping, following Papers~I--III. We first draw the sample size as a Poisson deviate with expectation equal to the observed number of TDEs in the standard sample. We then draw that many TDEs from the observed sample with replacement and compute the rate using Equation~\ref{eq:rate_combined}. The procedure is repeated $5\times10^{4}$ times, and the $1\sigma$ uncertainties are the $16^{\mathrm{th}}$ and $84^{\mathrm{th}}$ percentiles of the resulting rate distribution. For subsets or bins with $N \le 2$ TDEs, where the bootstrap is unreliable, we instead scale the $1\sigma$ Poisson confidence limits of \citet{Gehrels1986} to the measured rate (0.17--3.30 and 0.35--2.32 times the rate for $N=1$ and 2). Rates based on so few events are correspondingly uncertain and sensitive to individual objects. Uncertainties on the relative fractions use a joint bootstrap, in which each object's resampling weight enters both the subset and total rates, preserving their correlation. For $N \le 2$, the subset members instead receive Gamma-distributed weights that reproduce the \citet{Gehrels1986} limits, avoiding the spurious zero lower bound that arises when a lone object is dropped from a resample. Because these uncertainties are large and asymmetric, we quote rates as $\log_{10}(R)$ throughout. The resulting rates and fractions by spectral type are given in Table~\ref{tab:tde_rates}.

\begin{table}
\centering
\caption{Volumetric TDE rates and relative fractions by spectral type.
All fractions are relative to the total rate. Because each subtype rate
is a strict subset of the total, $R_i$ and $R_\mathrm{tot}$ are
correlated; we account for this with a joint bootstrap rather
than linear error propagation.}
\label{tab:tde_rates}
\renewcommand{\arraystretch}{1.7}
\setlength{\tabcolsep}{4.4pt}
\resizebox{\columnwidth}{!}{%
\begin{tabular}{l | c c c c}
\toprule
Sample & $z_\mathrm{med}$ & $N$ & $\log_{10}(R\,[\Runit])$ & Fraction [\%] \\
\midrule
TDE-H+He  & 0.033 & 15 & $2.31^{+0.15}_{-0.23}$  & $96.6^{+1.8}_{-3.0}$\\
TDE-H     & 0.045 & 4  & $0.77^{+0.20}_{-0.40}$  & $2.8^{+2.7}_{-1.7}$\\
TDE-He    & 0.080 & 2  & $0.06^{+0.37}_{-0.45}$ & $0.55^{+1.04}_{-0.36}$\\
\midrule
\textbf{Total} & \textbf{0.044} & \textbf{21} & $\mathbf{2.32^{+0.15}_{-0.23}}$ & \textbf{100} \\
\bottomrule
\end{tabular}%
}
\end{table}

For our full standard sample of 21 TDEs, we find a total volumetric rate of
\begin{equation}
    \log_{10}(R_{\rm tot}\,[\Runit]) = 2.32^{+0.15}_{-0.23},
    \label{eq:rtot}
\end{equation}
i.e., $R_{\rm tot} = 210 \pm 85~\Runit$. This total rate is effectively limited to $L_{g,\mathrm{peak}} \gtrsim 10^{42.2}$~erg~s$^{-1}$ ($M_{g,\mathrm{peak}} \lesssim -17$~mag), set by our faintest event, TDE~2023clx \citep[ASASSN-23bd;][]{Hoogendam2024b}, which alone contributes $\approx$27\% of $R_{\rm tot}$. From their sample of optical TDEs in ZTF, \citet{Yao2023} find $310^{+60}_{-100}$~yr$^{-1}$~Gpc$^{-3}$ for events with $L_{\rm BB} > 10^{43}$~erg~s$^{-1}$. Applying the same cut, which removes only TDE~2023clx from our standard sample, we find $\log_{10}(R_{L_{\rm BB}>10^{43}}\,[\Runit]) = 2.18^{+0.15}_{-0.23}$ ($R_{L_{\rm BB}>10^{43}} = 152 \pm 62$~$\Runit$), consistent with \citet{Yao2023} at 1.3$\sigma$, although the samples and selection functions differ.

To estimate a per-galaxy rate, we restrict to the 19 TDE hosts with $10^{9.5} \leq M_\star \leq 10^{11.5}\,M_\odot$. This excludes TDE~2023clx, with $M_\star = 10^{8.8}\,M_\odot$, and the off-nuclear TDE~2024tvd, for which we cannot make a robust stellar-mass estimate \citep[e.g.,][]{Patra2025}, giving $\log_{10}(R_{9.5-11.5}\,[\Runit]) = 2.18^{+0.15}_{-0.23}$. Dividing this by the number density of galaxies in the same mass range from the stellar mass function of \citet{Baldry2012} gives an average per-galaxy rate of 
\begin{equation}
    \log_{10}(r_{\rm gal}\,[\mathrm{yr^{-1}\,galaxy^{-1}}]) = -4.72^{+0.15}_{-0.23},
\end{equation}
i.e., $r_{\rm gal} = (1.9 \pm 0.8)\times10^{-5}$~yr$^{-1}$~galaxy$^{-1}$. This is broadly consistent with early estimates of the per-galaxy TDE rate from SDSS \citep{vanVelzen2014} and ASAS-SN \citep{Holoien2016a}. Our per-galaxy rate estimate is also within 1.3$\sigma$ of the $3.2^{+0.8}_{-0.6}\times10^{-5}$~yr$^{-1}$~galaxy$^{-1}$ that \citet{Yao2023} find for galaxies with $M_{\rm gal} \approx 10^{10}\,M_\odot$, although ours is an average over the given mass range rather than a value at fixed galaxy mass.

\subsection{Off-Nuclear and Recurrent TDE Rates} \label{subsec:offnuc_rec}

Among the standard sample of TDEs is TDE 2024tvd \citep[ZTF22aaigqsr;][]{Yao2025a}, one of the few optically-selected off-nuclear TDEs \citep{Yao2025a, Stein2026, Li2026b}. From this event, we compute the rate of off-nuclear TDEs to be
\begin{equation}
    \log_{10}(R_{\rm off\text{-}nuclear}\,[\Runit]) = 0.39^{+0.52}_{-0.76}.
\end{equation}
This corresponds to $1.2^{+3.4}_{-1.0}\%$ of the total volumetric TDE rate, with uncertainties from a joint bootstrap. This is consistent with the conservative fractional range of $0.2\mbox{--}9.6$\% estimated by \citet{Stein2026} from the ZTF Bright Transient Survey \citep{Fremling2020, Perley2020}.

Our standard sample also includes four recurrent TDEs (rTDEs), each with two flares as of 2024 December 31: TDE 2019azh \citep[ASASSN-19dj;][]{Hinkle2021a, Yao2026a}, TDE 2018fyk \citep[ASASSN-18ul;][]{Wevers2023a, Pasham2024b}, TDE 2022dbl \citep[ASASSN-22ci;][]{Lin2024, Hinkle2024c, Makrygianni2025, Hinkle2026}, and TDE 2024pvu \citep[ZTF18abxnvoz;][]{Yao2026a}. These events are plausibly powered by repeating partial TDEs \citep[e.g.,][]{Yao2026a, Hinkle2026}. Counting each rTDE once, using a single-flare template based on its first flare in ASAS-SN, we measure the volumetric rate of rTDEs to be
\begin{equation} \label{eq:R_rec}
    \log_{10}(R_{\rm recurrent}\,[\Runit]) = 1.26^{+0.23}_{-0.62},
\end{equation}
which corresponds to $9^{+8}_{-6}$\% of all TDEs. If we include ASASSN-14ko \citep[e.g.,][]{Payne2021, Payne2022, Payne2023}, then the total TDE rate becomes $\log_{10}(R_{\rm tot}\,[\Runit]) = 2.34^{+0.14}_{-0.21}$ ($R_{\rm tot} = 220 \pm 85$~$\Runit$), and the rTDE rate is $\log_{10}(R_{\rm recurrent}\,[\Runit]) = 1.45^{+0.20}_{-0.35}$, increasing the rTDE fraction to $13^{+10}_{-7}$\%.  Because recurrence can only be established when multiple flares are observed, this represents a lower limit on the intrinsic rTDE rate. Our central estimates ($\sim$$9\mbox{--}13\%$) are systematically lower than the $25\%\mbox{--}60\%$ fractional rate derived by \citet{Yao2026a} from a combination of Catalina Real-Time Transient Survey \citep[CRTS;][]{Drake2009} and ZTF observations, but demonstrate that rTDEs are a non-negligible fraction of all TDEs.

Of the four rTDEs in our standard sample, only TDE~2018fyk (ASASSN-18ul) and TDE~2022dbl (ASASSN-22ci) had a second optical flare during the 2014--2024 ASAS-SN baseline; the additional flares of TDE~2019azh (ASASSN-19dj) and TDE~2024pvu (ZTF18abxnvoz) occurred before ASAS-SN began survey operations. The second flare of TDE~2018fyk peaked fainter than $m_{g,\rm peak} = 18$~mag and therefore does not pass the cuts of our standard sample. This leaves a single additional flare, the 2024 re-brightening of TDE~2022dbl \citep{Lin2024, Hinkle2024c}, which peaked at $m_{g,\rm peak} = 17.8$~mag. Treating this flare exactly as any other TDE in our sample, with $F_1$, $F_2$, and $F_3$ evaluated at its own peak magnitude and year, and counting each flare as an independent event, we find
\begin{equation}
    \log_{10}(R_{\rm recurrent,\,flares}\,[\Runit]) = 1.91^{+0.25}_{-0.72},
\end{equation}
which corresponds to $30^{+19}_{-23}\%$ of all TDE flares in our sample, counting the TDE~2022dbl re-brightening as an additional event. This is a factor of $\sim$4 higher than the rate obtained when each rTDE is counted once (Equation~\ref{eq:R_rec}), because the fainter second flare of TDE~2022dbl ($M_g = -17.6$~mag) is only detectable out to $z \approx 0.03$ and therefore carries a large weight, accounting for about three-quarters of $R_{\rm recurrent,\,flares}$.

\section{Optical Luminosity Functions} \label{sec:gband_LF}

\begin{figure*}
    \centering
    \includegraphics[width=\textwidth]{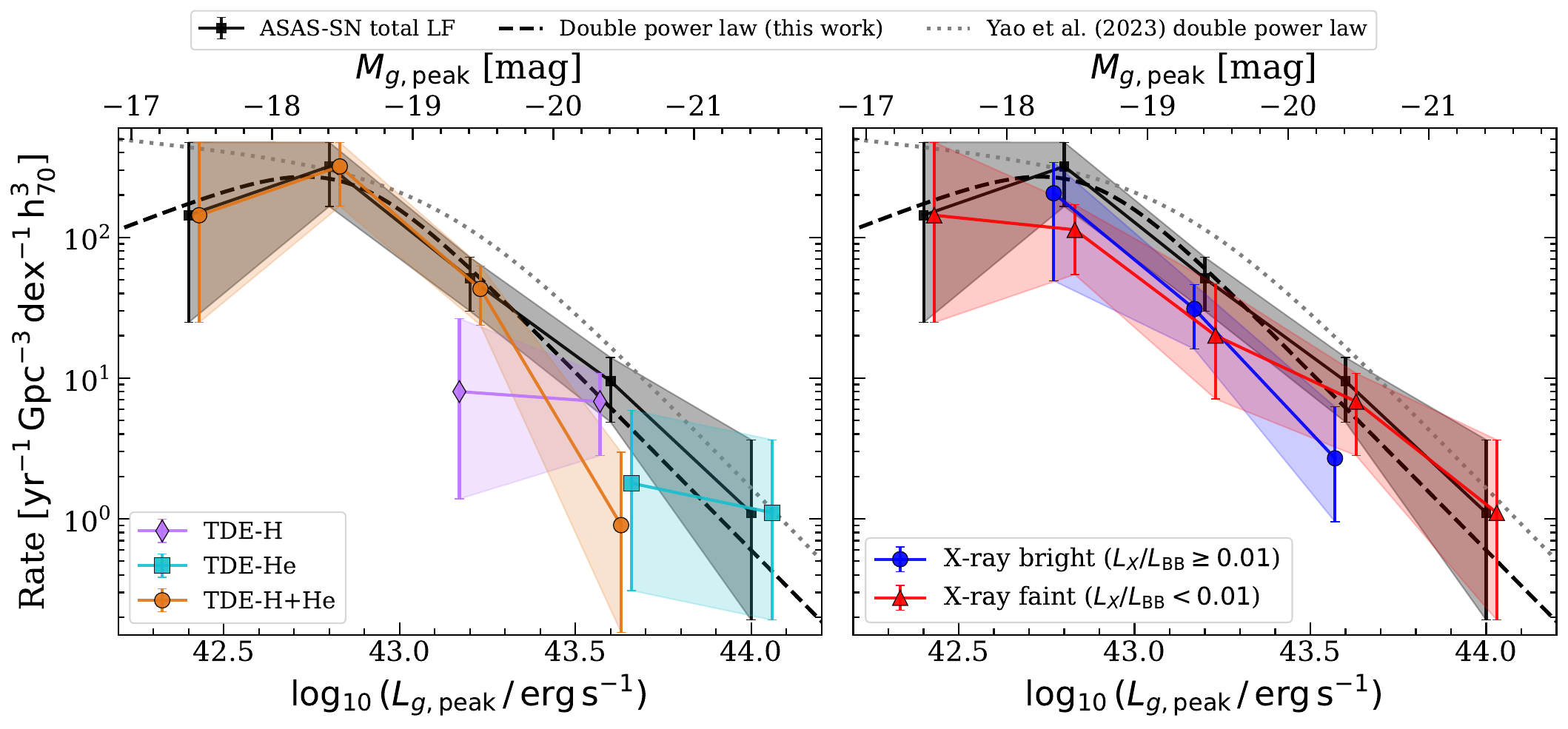}
    \caption{ASAS-SN $g$-band TDE LF. The TDE rate increases towards lower luminosity, becoming shallower or possibly plateauing below $\sim$$10^{42.7}$ erg s$^{-1}$. The black dashed line shows a double power-law fit to our TDE LF. The double power-law fit to the ZTF $g$-band TDE LF of \citet{Yao2023} is shown as the dotted gray line for comparison. \textit{Left Panel:} The $g$-band LF, colored by the optical spectroscopic type of the TDE. Shown are TDE-H (purple), TDE-He (cyan), and TDE-H+He (orange). \textit{Right Panel:} The $g$-band LF, colored by the X-ray-to-blackbody luminosity ratio of the TDE. TDEs with $L_\mathrm{X}/L_\mathrm{BB} \geq 0.01$ are shown in blue and events with $L_\mathrm{X}/L_\mathrm{BB} < 0.01$ are shown in red. In both panels, the total rate is given by the black line. Subset points are offset slightly from the bin centers for clarity.}
    \label{fig:gband_LF}
\end{figure*}

We construct the $g$-band LFs in terms of the peak rest-frame $g$-band luminosity, which we define as $L_g \equiv \nu L_\nu$ at $\nu \approx 6.3\times10^{14}$~Hz, and compute from the peak absolute magnitude as $\log_{10}(L_g/\mathrm{erg\,s^{-1}}) = 35.44 - 0.4\,M_{g,\mathrm{peak}}$, where the constant follows from the AB zero point (3631~Jy) at 10~pc. 

The sections that follow leverage the design of ASAS-SN and our well-calibrated knowledge of its systematics to examine various comparisons of the rate of optical TDEs as a function of key TDE parameters of interest. First, we determine the peak $g$-band luminosity function (this Section) both for the full sample and as a function of optical spectroscopic type (Section \ref{subsec:spec_types}) and X-ray luminosity (Section \ref{subsec:xray_lum}). Next, we compare the TDE rate to properties at peak light inferred through blackbody modeling of the TDE UV/optical SED (Section \ref{sec:BB_rates}). Finally, we investigate how the TDE rate varies with host-galaxy properties (Section \ref{sec:hostgal_rates}), such as the QBS or PSB nature of the host (Section \ref{subsec:host_qbs_psb}), a recent burst component (Section \ref{subsec:burst_props}), the bulge S\'{e}rsic index (Section \ref{subsec:host_sersic}), and the SMBH mass (Section \ref{subsec:smbh_mass_func}).

Figure~\ref{fig:gband_LF} presents the $g$-band LF of TDEs in our sample; the binned rates for this and the later figures are given in Appendix~\ref{app:binned_rates}. Over the luminosity range probed by our sample, the $g$-band LF declines toward higher peak optical luminosities, consistent with previous results \citep{vanVelzen2018, Lin2022, Charalampopoulos2023, Yao2023}. The shape is similar to the LF produced by ZTF \citep{Yao2023}, although our rates are lower by a factor of $\sim$2 at $L_g \gtrsim 10^{43}$~erg~s$^{-1}$, consistent with the ratio of the integrated rates for $L_{\rm BB} > 10^{43}$~erg~s$^{-1}$, $R_{\rm ZTF}/R_{\rm ASAS\text{-}SN} = 2.0^{+1.4}_{-0.8}$ (Section~\ref{subsec:vol_rates}). \citet{Yao2023} measured a break in the rates at $\log_{10}(L_g/\mathrm{erg\,s^{-1}}) = 43.1$ with a rate normalization of $287$ yr$^{-1}$ Gpc$^{-3}$ dex$^{-1}$. Employing the same model, 
\begin{equation}
    \phi(L_g) \equiv \frac{dR}{d\log_{10} L_g} = \dot{N}_0 \left[ \left(\frac{L_g}{L_{\rm bk}}\right)^{\gamma_1} + \left(\frac{L_g}{L_{\rm bk}}\right)^{\gamma_2} \right]^{-1},
    \label{eq:dpl}
\end{equation}
where $L_{\rm bk}$ is the break luminosity and $\phi \propto L_g^{-\gamma_1}$ ($L_g^{-\gamma_2}$) well below (above) the break, we fit our LF and estimate the parameters from $10^{4}$ bootstrap resamplings of the sample, adopting the median and 16th--84th percentiles of each parameter. Because our faint end contains only a single event, we retain only the bootstrap fits with $\gamma_1 > -10$, excluding unphysically steep solutions. We find a break luminosity of $\log_{10}(L_g/\mathrm{erg\,s^{-1}}) = 42.9^{+0.7}_{-0.2}$, a normalization of $\dot{N}_0 = 480^{+430}_{-440}$ yr$^{-1}$ Gpc$^{-3}$ dex$^{-1}$ $h_{70}^{3}$, a bright-end slope of $\gamma_2 = 2.5^{+1.8}_{-0.5}$, and a faint-end slope of $\gamma_1 = -0.9^{+1.7}_{-2.1}$, all consistent with \citet{Yao2023} within the respective uncertainties. A larger sample is needed to test whether the faint end of the $g$-band LF turns over or continues to rise.

Exotic accretion-powered events with high peak luminosities such as luminous ANTs \citep{Wiseman2025} and extreme nuclear transients \citep[ENTs;][]{Hinkle2025} are broadly consistent with extrapolations of the high-end of the TDE LF, although this does not necessarily imply that ANTs and ENTs are drawn from the same physical population as typical UV/optical TDEs. The rate of luminous ANTs was estimated from a ZTF-selected sample to be $\gtrsim 3\times10^{-2}$ yr$^{-1}$ Gpc$^{-3}$ \citep[][]{Wiseman2025}. This rate occurs at $\log_{10}(L_g/\mathrm{erg\,s^{-1}}) = 44.6$, near the median ANT peak luminosity \citep[e.g.,][]{Neustadt2020, Wiseman2025}. The ENT rate was loosely constrained to be $\gtrsim 1\times10^{-3}$ yr$^{-1}$ Gpc$^{-3}$ \citep[][]{Hinkle2025} from a sample of only 3 events. An extrapolation of our $g$-band LF yields this rate at $\log_{10}(L_g/\mathrm{erg\,s^{-1}}) = 45.1$, in reasonable agreement with the confirmed ENTs. Future rate comparisons will need to account for the poorly constrained completeness for these events, with the recent discovery of several ENT candidates at lower redshifts \citep[][]{Quintin2026} suggesting that the true rate may be higher. The indications of pre-flare AGN activity, if it is related to the production of ANT and ENT-like flares, could decrease the rates by roughly a factor of the AGN fraction \citep[$\sim$5\%;][]{Ho2008, Lacerda2020}.

\subsection{Optical Spectroscopic Types} \label{subsec:spec_types}
 
Next, we examine the rates of the different optical spectroscopic types of TDEs: TDE-H, TDE-He, TDE-H+He, and TDE-Featureless \citep[e.g.,][]{Leloudas2019, vanVelzen2021, Hammerstein2023}. Here we adopt the near-peak spectroscopic type, although some TDEs show evolution in their spectra \citep[e.g.,][]{Nicholl2019, Short2020, Hinkle2021a}. Observational studies have found differences in the physical properties of different TDE spectroscopic types, including the larger blackbody radii of TDE-H events \citep{Hinkle2020a, vanVelzen2021, Nicholl2022}, the high temperatures of TDE-He and TDE-Featureless events \citep{vanVelzen2021, Hammerstein2023}, and a preference for lower (higher) mass SMBHs for TDE-H (TDE-Featureless) events \citep[][]{Nicholl2022, Hammerstein2023, Yao2026b}. 

Theoretical studies indicate that high-luminosity and compact photospheres are highly ionized and can suppress prominent optical line features, producing featureless spectra \citep[e.g.,][]{Aspegren2026, Thomsen2026}. Consistent with the observations, these studies also predict that the strongest emission lines vary with luminosity and photospheric radius, with H lines present for sufficiently extended photospheres and progressively stronger He II emission in more compact photospheres \citep{Roth2016, Aspegren2026, Thomsen2026}. For Eddington-limited TDE emission, the observed SMBH mass trend is also consistent with He-dominated or featureless spectra corresponding to higher luminosity events.
 
As shown in Table \ref{tab:tde_rates}, TDE-H+He events heavily dominate the rate of nearby TDEs, comprising $96.6^{+1.8}_{-3.0}$\% of the volumetric rate. The TDE-H and TDE-He spectroscopic types occur at $2.8^{+2.7}_{-1.7}$\% and $0.55^{+1.04}_{-0.36}$\% of the total optical TDE rate, respectively. The extremely high fraction of events showing both strong H and He emission is consistent with a comparison of the typical TDE peak blackbody luminosities and radii to parameters expected to produce such spectra \citep[][cf. their Figure 3]{Aspegren2026} and broadly agrees with the predictions of \citet[][cf. their Figures 7 and 8]{Roth2016} and \citet[][cf. their Figure 2d]{Thomsen2026}.

The TDE-Featureless events comprise two sub-populations \citep[e.g.,][]{Yao2026b}, one at high luminosities where over-ionization is likely \citep{Aspegren2026} and another with lower luminosities but more compact photospheres \citep{Yao2026b}. ASAS-SN has not discovered or recovered an unambiguous TDE-Featureless event. We place a $95\%$ confidence upper limit on the rate of luminous ($-20$ to $-23$~mag) featureless TDEs of $\log_{10}(R\,[\Runit]) < -0.04$, or $<$$0.5\%$ of the total rate, assuming Poisson statistics for zero detected events. To compute the selection efficiency, we inject $10^6$ events uniformly in comoving volume out to $z = 0.12$ with peak magnitudes between $-20$ and $-23$~mag, and recover them using the completeness curves built from the light curve of the luminous TDE-Featureless event TDE~2020qhs \citep{Hammerstein2023} for each survey year.

The enigmatic transient ASASSN-15lh, which shows no clear emission features, has been interpreted both as a luminous TDE candidate \citep[e.g.,][]{Leloudas2016} and as a superluminous supernova \citep[e.g.,][]{Dong2016}. Here, assuming that it is a featureless TDE, we conservatively use ASASSN-15lh as a constraint on the potential rate of luminous TDE-Featureless events. At $z = 0.2326$, it lies beyond the $z < 0.12$ volume of our standard sample, so we compute its rate using the $1/V_{\rm max}$ method, summing $V_{\rm max}$ over the two survey eras, with $V_{\rm max}$ set by the redshift to which ASAS-SN could have detected it in each era ($z_{\rm lim} = 0.25$ in the $V$-band era and $0.41$ in the $g$-band era). We find $\log_{10}(R_{\rm 15lh}\,[\Runit]) = -1.92^{+0.52}_{-0.76}$, or $0.006^{+0.017}_{-0.005}\,\%$ of the total volumetric TDE rate, well below the upper limit placed on luminous featureless TDEs.

At the other end of the luminosity function, ASASSN-20hx \citep[AT~2020ohl;][]{Hinkle2022} has been claimed to be a low-luminosity featureless TDE \citep{Yao2026b}. It has been alternatively classified as an ANT \citep{Hinkle2022, Roy2024, Hinkle2024b}. The continued optical excess above host-galaxy levels and ongoing years-long re-brightening\footnote{\url{http://asas-sn.ifa.hawaii.edu/skypatrol/objects/223338499160}} are highly atypical of TDEs, making it uncertain whether it is a bona fide TDE-Featureless. Nevertheless, assuming ASASSN-20hx is a low-luminosity TDE-Featureless event, we derive an approximate constraint on the rate of low-luminosity TDE-Featureless events within ASAS-SN of $\log_{10}(R\,[\Runit]) = 1.31^{+0.52}_{-0.76}$, or $9^{+19}_{-7} \%$ of the total volumetric TDE rate.
 
In the left panel of Figure \ref{fig:gband_LF}, we see that TDE-H+He events are dominant for all $g$-band luminosities below $\approx$$2\times10^{43}$ erg s$^{-1}$. By $\approx$$4\times10^{43}$ erg s$^{-1}$, the rates of TDE-H, TDE-H+He, and TDE-He events are similar, and only the TDE-He events in our sample have higher $g$-band luminosities. The low rates and high luminosities of TDE-He events, and the absence of luminous TDE-Featureless events from our sample, are generally consistent with the preference of such TDEs to occur on more massive SMBHs.

\subsection{X-ray Luminosity} \label{subsec:xray_lum}

\begin{figure*}
    \centering
    \includegraphics[width=\textwidth]{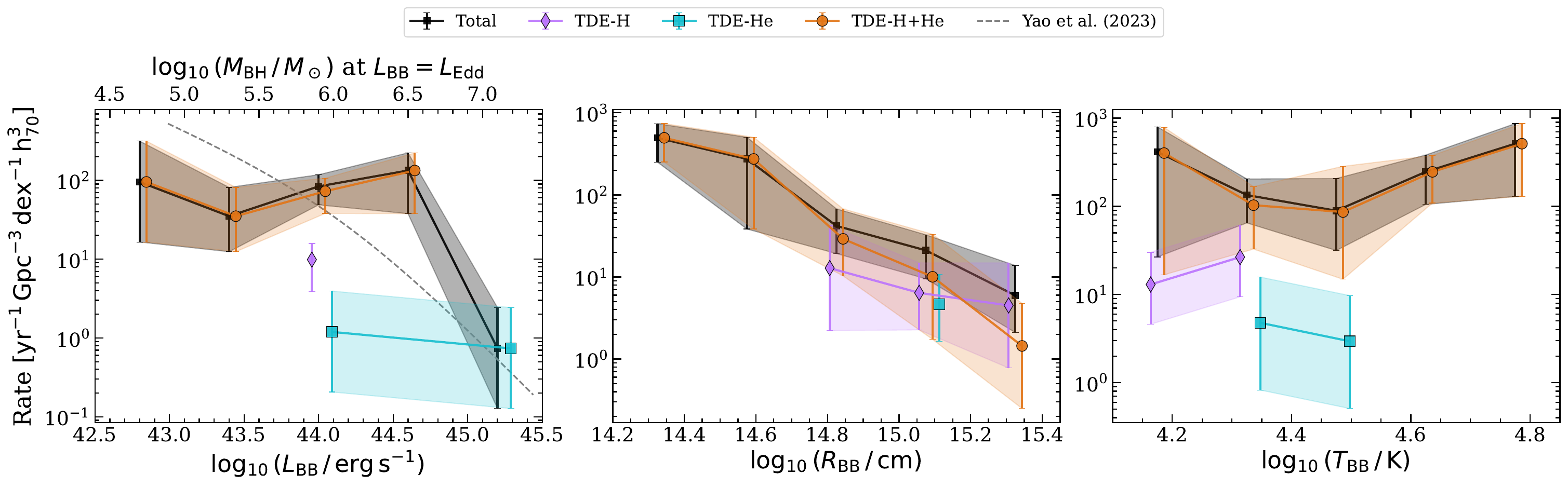}
    \caption{Optical TDE rates compared to the peak blackbody luminosity (left panel), blackbody radius at peak (middle panel), and blackbody temperature at peak (right panel). The rates are colored by the optical spectroscopic type of the TDE. Shown are TDE-H (purple), TDE-He (cyan), and TDE-H+He (orange). The total rate is shown in black. In the left panel, the gray dashed line shows the double power-law fit to the blackbody LF of \citet{Yao2023}, and the top axis gives the SMBH mass for which the blackbody luminosity equals the Eddington luminosity (Section~\ref{sec:BB_rates}). Subset points are offset slightly from the bin centers for clarity.}
    \label{fig:LF_BB_props}
\end{figure*}
 
While the TDEs in our sample are, by definition, optically-selected, they all have Swift XRT \citep{Burrows2005} coverage. Therefore, we can also examine the $g$-band LF with respect to the X-ray luminosity of the TDEs. This comparison is shown in the right panel of Figure \ref{fig:gband_LF}, with two bins separating the TDEs at an X-ray-to-blackbody luminosity ratio of $L_\mathrm{X}/L_\mathrm{BB} = 0.01$, approximately the median of our sample. The normalization and shape of our $g$-band LF is broadly similar to TDE X-ray LFs \citep[e.g.,][]{Auchettl2018, Sazonov2021, Guolo2024, Grotova2025}. Given the heterogeneous coverage, $L_\mathrm{X}$ should be viewed as a lower limit on the intrinsic maximum X-ray luminosity for many events. We therefore use $L_\mathrm{X}/L_\mathrm{BB}$ as an approximate X-ray-brightness indicator, although the upper limits for all TDEs without detected X-ray emission lie firmly in the $L_\mathrm{X}/L_\mathrm{BB} < 0.01$ bin. We additionally caveat that the observed X-ray luminosity can vary strongly as a function of time, with some TDEs showing a significant X-ray brightening several months after peak optical light \citep[e.g.,][]{Auchettl2017, Gezari2017, Holoien2018, Hinkle2021a}.
 
The rates of X-ray-faint and X-ray-bright optical TDEs are nearly identical. A qualitatively similar result is recovered when the sample is split at $L_\mathrm{X} = 10^{42}$ erg s$^{-1}$ \citep[cf.][approximately the median X-ray luminosity of our sample]{Hammerstein2023}. This suggests that the physics setting the multi-wavelength SED appearance, as traced here by the volumetric rate of X-ray-bright and X-ray-faint optical TDEs, is likely a viewing angle effect \citep[e.g.,][]{Auchettl2017, Dai2018a, Thomsen2022}. In this picture, TDEs are powered by promptly-formed accretion disks, and the broadband appearance depends on the optical depth of a reprocessing outflow along the line of sight to an observer. Observers whose lines of sight pass through an optically thick reprocessing layer preferentially observe UV/optical-dominated emission, whereas observers viewing the inner accretion flow without significant obscuration can detect stronger X-ray emission. This event-wise constraint is reminiscent of the global similarity of the volumetric rates of X-ray-selected and optically-selected TDEs \citep[e.g.,][]{Auchettl2018, Sazonov2021, Yao2023, Guolo2024, Grotova2025}.

\section{Rates vs. Blackbody Properties} \label{sec:BB_rates}
 
Despite the optical luminosity being a key direct observable, optically-selected TDEs emit the majority of their radiation in the UV, with SEDs well-described by single-temperature blackbodies \citep[e.g.,][]{Gezari2012, Holoien2016a, Hinkle2020a, vanVelzen2021}. We therefore consider the dependence of optical TDE rates on their blackbody properties at peak light (see Section \ref{subsec:tde_properties} for more details) in Figure \ref{fig:LF_BB_props}.
 
Similar to previous results \citep{Lin2022, Yao2023}, we find that the decline in TDE rate towards higher blackbody luminosities is shallower than the $g$-band LF discussed in Section \ref{sec:gband_LF}. Unlike the blackbody LFs produced from ZTF data \citep{Lin2022, Yao2023}, a single power-law is not a good fit to the data. The integrated rates are consistent (Section~\ref{subsec:vol_rates}), but our blackbody LF is flatter for $L_\mathrm{BB} \lesssim 10^{44.5}$ erg s$^{-1}$. While there are likely small differences in the methodology of estimating blackbody parameters, we suspect that this primarily reflects the limited number of low-luminosity events in our sample. Our two lowest-luminosity bins have only 3 sources, as compared to $\approx$8 TDEs in the two lowest-luminosity bins for \citet{Yao2023}. Many of these low-luminosity TDEs peak at apparent magnitudes similar to or fainter than the ASAS-SN limiting magnitude and often have faster light curve evolution than more luminous TDEs \citep[e.g.,][]{Charalampopoulos2023, Hinkle2023, Hoogendam2024b}, making them difficult to detect. As compared to ASAS-SN, the larger volume probed by ZTF makes it more sensitive to these intrinsically faint, yet intrinsically common, events at modest redshift. Regardless, the blackbody LF at $L_\mathrm{BB} \lesssim 10^{44.5}$ erg s$^{-1}$ remains consistent with Eddington-limited luminosities (see Section \ref{subsec:smbh_mass_func}). The top axis of the left panel of Figure~\ref{fig:LF_BB_props} gives the SMBH mass for which $L_\mathrm{BB}$ equals the Eddington luminosity, $L_\mathrm{Edd} = 1.26\times10^{38}\,(M_\mathrm{BH}/M_\odot)$ erg s$^{-1}$. We define the UV/optical Eddington ratio of each TDE as $\lambda_\mathrm{Edd} \equiv L_\mathrm{BB}/L_\mathrm{Edd}$, using the peak blackbody luminosity and the SMBH mass from Section~\ref{subsec:smbh_masses}.

Although the conversion between mass fallback and luminosity is complicated, the maximum accretion rate of a TDE is set primarily by the mass of the disrupted star \citep[e.g.,][]{Kochanek2016b, Mockler2019, Bandopadhyay2024a}. The low rates of the highest-luminosity TDEs ($\sim$1 yr$^{-1}$ Gpc$^{-3}$ $h_{70}^{3}$) are suggestive of the disruption of massive stars, which should be intrinsically rare due to the stellar initial mass function and short stellar lifetimes.

We recover the known trend for TDE-H events to occur at lower blackbody temperatures \citep[e.g.,][]{Hinkle2020a, vanVelzen2021, Hammerstein2023}. There is no significant relationship between the TDE rate and blackbody temperature, with an approximately equal volumetric rate of TDEs for any observed UV/optical blackbody temperature. 

\begin{figure*}
    \centering
    \includegraphics[width=\textwidth]{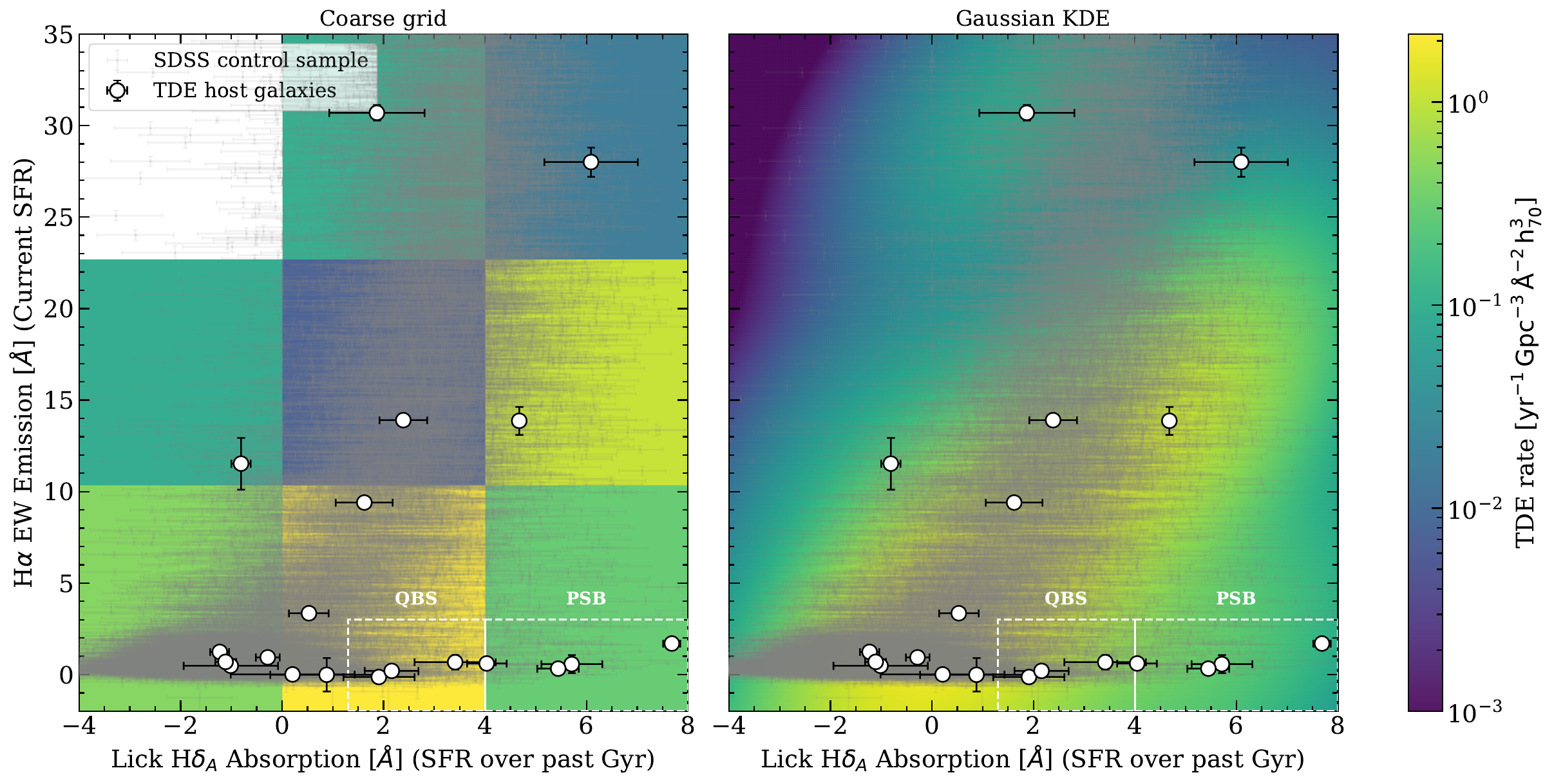}
    \caption{H$\alpha$ emission line EW, which traces current star formation, as compared to the Lick H$\delta_A$ absorption index, which traces star formation in the past Gyr. The TDE host galaxies in our sample are shown in white, and the gray points are the SDSS control sample. Volumetric TDE rates are shown in coarse bins (left panel) and as a Gaussian-smoothed KDE (right panel).}
    \label{fig:Halpha_Hdelta_rates}
\end{figure*}

Like previous studies \citep{vanVelzen2018, Yao2023}, the optical TDE rate declines monotonically with increasing blackbody radius. The low rate at blackbody radii of $\sim$2$\times 10^{15}$ cm broadly agrees with the break radius found by \citet{Yao2023}. Their interpretation that the presence of a maximum radius at this scale supports a limit imposed by the semi-major axis distance of the most bound debris is consistent with our observations and the rate suppression seen at the high-mass end of our TDE BH mass function (see Section \ref{subsec:smbh_mass_func}).

\section{Rates vs. Host-Galaxy Properties} \label{sec:hostgal_rates}

Transient host galaxies likely hold information relevant to understanding the physical origins of the transients themselves. Although the basic physical picture of a TDE is well understood, large open questions remain about the rate at which stars are placed on the low angular momentum orbits that ultimately lead to their destruction in a TDE. Here we discuss the rate of optical TDEs as a function of various properties of their host galaxies, including the per-galaxy rate enhancement in QBS and PSB galaxies, the presence of a recent burst of star formation, the host-galaxy light profile, and the mass of the host-galaxy SMBH.
 
We exclude TDE~2024tvd from the host-galaxy measurements that follow, since properties measured at the host nucleus do not describe the environment of its disruption. This leaves 20 of the 21 TDEs in our standard sample. TDE~2024tvd is retained in Section~\ref{subsec:smbh_mass_func}, as its SMBH mass is estimated from the TDE itself rather than from host-galaxy scalings.

\subsection{Rate Enhancements in QBS and PSB Galaxies} \label{subsec:host_qbs_psb}

It has long been recognized that many TDEs occur in galaxies with little current star formation, but evidence for substantial recent (within $\sim$1 Gyr) star formation \citep[e.g.,][]{Arcavi2014, French2016, French2018, LawSmith2017, Graur2018a}. It is now well-established that the per-galaxy TDE rate is enhanced in QBS and PSB galaxies, at levels of $\sim$$10\mbox{--}100\times$ the average rate \citep[e.g.,][]{French2020b}, with galaxies at more extreme Lick H$\delta_A$ index typically having higher rate enhancements.
 
Figure \ref{fig:Halpha_Hdelta_rates} compares H$\alpha$ emission EW, tracing current star formation, to the Lick H$\delta_A$ absorption index, tracing star formation within the past Gyr, colored by the volumetric TDE rate. The KDE is a two-dimensional Gaussian kernel density estimate of the host positions in this plane, with each TDE entering with its completeness weight $w_i$. We use the \texttt{gaussian\_kde} implementation in \texttt{SciPy} \citep{Virtanen2020} with Scott's rule for the bandwidth \citep{Scott2015}: the kernel covariance is the weighted sample covariance of the host positions scaled by $n_{\mathrm{eff}}^{-1/3}$, where $n_{\mathrm{eff}} = (\sum_i w_i)^2 / \sum_i w_i^2 \approx 6.0$ is the effective number of hosts given the unequal weights.   This corresponds to kernel widths of $\approx$1.8\,\AA\ in Lick H$\delta_A$ and $\approx$5.5\,\AA\ in H$\alpha$ EW. The resulting surface is renormalized to integrate to the total volumetric rate of the sample, so that both panels show a rate per unit area of the plane and share a single color scale. Both for the discrete bins and the smoothed KDE, the volumetric TDE rate largely follows the underlying galaxy distribution. This independently confirms that QBS and PSB galaxies do not dominate the volumetric TDE rate, despite their significant per-galaxy rate enhancements. Volumetrically, the typical TDE host is a weakly star-forming galaxy, consistent with known TDE hosts \citep[e.g.,][]{French2016, Graur2018a, French2020b}.
 
The over-representation of TDEs in the lower-right-hand corner, where QBS and PSB galaxies reside, is clear in Figure \ref{fig:Halpha_Hdelta_rates}, so we quantify the rate enhancements. Following \citet{French2016}, we define QBS galaxies by H$\alpha$ EW $< 3$~\AA\ and $1.3~\text{\AA} \leq \mathrm{Lick} \ \mathrm{H}\delta_A < 4$~\AA, and PSB galaxies as those with H$\alpha$ EW $< 3$~\AA\ and $\mathrm{Lick} \ \mathrm{H}\delta_A \geq 4$~\AA. Of the 20 hosts, three are classified as QBS and four as PSB.

\begin{figure*}
    \centering
    \includegraphics[width=\textwidth]{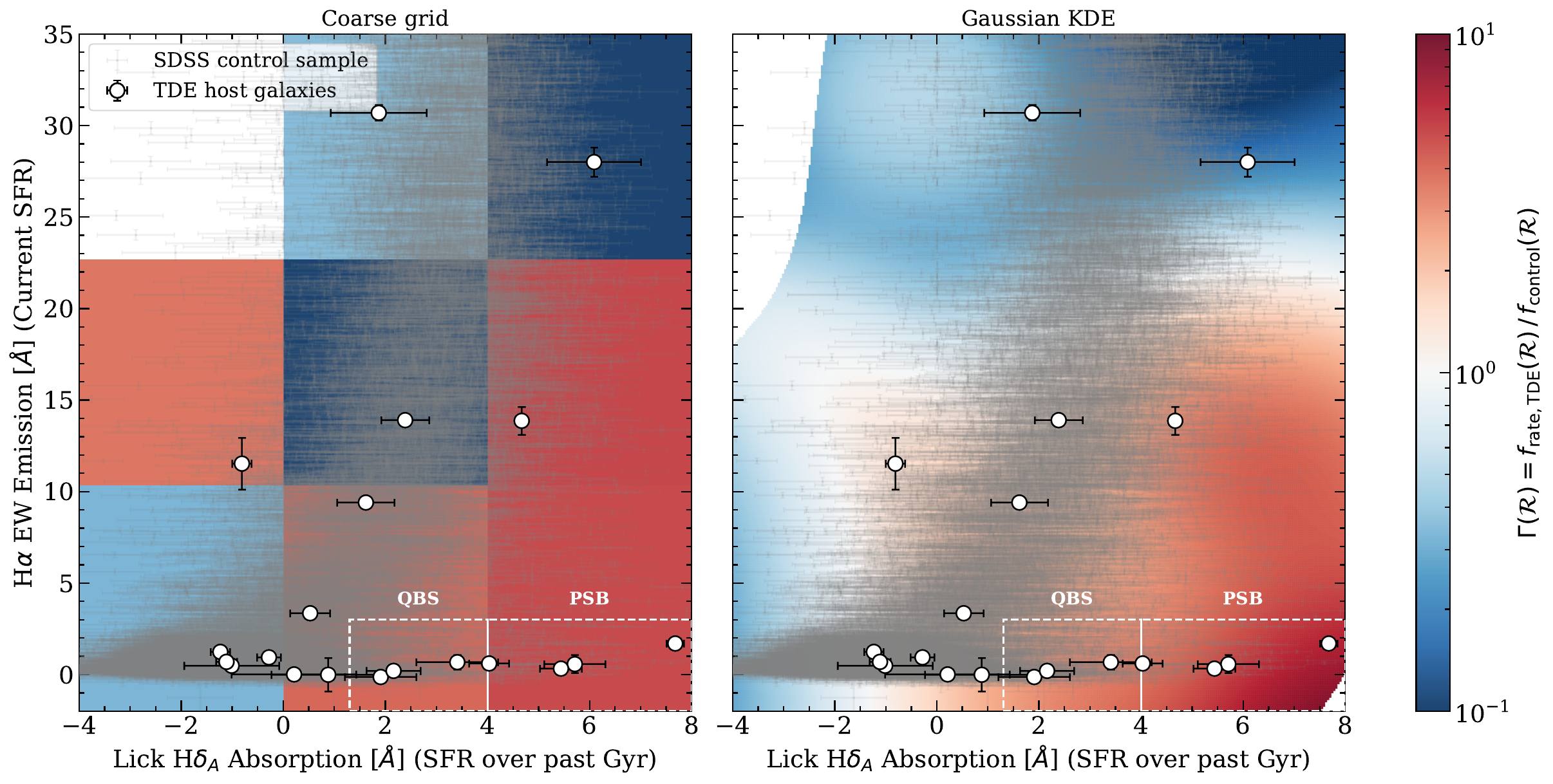}
    \caption{The same as Figure \ref{fig:Halpha_Hdelta_rates}, except that the colors refer to the TDE rate enhancement (see Equation~\ref{eq:enhancement}). The rate enhancements are shown in coarse bins (left panel) and as a Gaussian-smoothed KDE (right panel).}
    \label{fig:Halpha_Hdelta_enhancement}
\end{figure*}
 
Because our TDE sample is completeness-corrected while the control sample is not, the two cannot be compared directly by number. Each TDE carries the statistical weight $w_i = (F_{1,i}\,F_{2,i}\,F_{3,i})^{-1}$ defined in Section~\ref{sec:rates}. We therefore assign a corresponding weight to each control galaxy. For control galaxy $g$ in cell $c$ of redshift and stellar mass parameter space,
\begin{equation}
    \omega_g = \frac{W_c}{N_c}, \qquad \text{where} \quad
    W_c = \sum_{i \in c} w_i,
    \label{eq:control_weight}
\end{equation}
with $N_c$ the number of control galaxies in cell $c$. This distributes the total TDE weight in each cell uniformly among that cell's control galaxies, and guarantees $\sum_g \omega_g = \sum_i w_i$ by construction. For cells containing a single TDE host, Equation~\ref{eq:control_weight} reduces to $\omega_g = w_i/950$.
 
The TDE rate and control galaxy fractions in a region $\mathcal{R}$ of the H$\alpha$--H$\delta_A$ plane are then
\begin{equation}
    f_\mathrm{rate,TDE}(\mathcal{R}) =
    \frac{\sum_{i \in \mathcal{R}} w_i}{\sum_i w_i}
    \label{eq:f_rate}
\end{equation}
and
\begin{equation}
    f_\mathrm{control}(\mathcal{R}) =
    \frac{\sum_{g \in \mathcal{R}} \omega_g}{\sum_g \omega_g}
    \label{eq:f_con}
\end{equation}
with their ratio being the rate enhancement
\begin{equation}
    \Gamma(\mathcal{R}) =
    \frac{f_\mathrm{rate,TDE}(\mathcal{R})}{f_\mathrm{control}(\mathcal{R})}.
    \label{eq:enhancement}
\end{equation}
\noindent $\Gamma = 1$ indicates that the estimate of the fraction of volumetric TDE rate in region $\mathcal{R}$ is exactly proportional to the weighted fraction of galaxies in that region, at matched stellar mass and redshift; $\Gamma > 1$ indicates overrepresentation. This rate enhancement is shown in Figure~\ref{fig:Halpha_Hdelta_enhancement}, and Table~\ref{tab:enhancement_all} gives its value in every region. For the KDE panel, the TDE and control distributions are smoothed with the same fixed Gaussian kernel used for Figure~\ref{fig:Halpha_Hdelta_rates}, before taking their ratio. We estimate uncertainties with the same joint Poisson bootstrap used in Section~\ref{sec:rates}, recomputing Equations~\ref{eq:control_weight}--\ref{eq:enhancement} in full for each realization so the numerator and denominator fluctuate coherently. We report the 16th and 84th percentiles as the uncertainties.

\begin{table}
\centering
\caption{Volumetric TDE rate enhancement factors $\Gamma$ in the
H$\alpha$ EW--Lick H$\delta_A$ plane, relative to a stellar-mass- and redshift-matched SDSS control sample. $\Gamma = 1$ corresponds to no enhancement, i.e.\ a volumetric TDE rate exactly proportional to the abundance of such galaxies at matched $M_\star$ and $z$. Uncertainties are the 16th and 84th percentiles of the joint Poisson bootstrap realizations. The right-hand column excludes TDE 2023clx, which carries over a quarter of the total sample weight but whose host lies in neither region.}
\label{tab:tde_enhancement}
\renewcommand{\arraystretch}{1.6}
\setlength{\tabcolsep}{4.4pt}
\begin{tabular}{l | c c c}
\toprule
Population & $N_\mathrm{TDE}$ & $\Gamma$ (Full sample) & $\Gamma$ (Excl.\ TDE 2023clx) \\
\midrule
PSB & 4 & $15^{+16}_{-11}$ & $27^{+18}_{-18}$\\
QBS & 3 & $5^{+5}_{-4}$   & $6^{+6}_{-5}$\\
\midrule
PSB+QBS & 7 & $7^{+5}_{-3}$ & $10^{+6}_{-5}$ \\
\bottomrule
\end{tabular}
\end{table}

\begin{figure*}
    \centering
    \includegraphics[width=\textwidth]{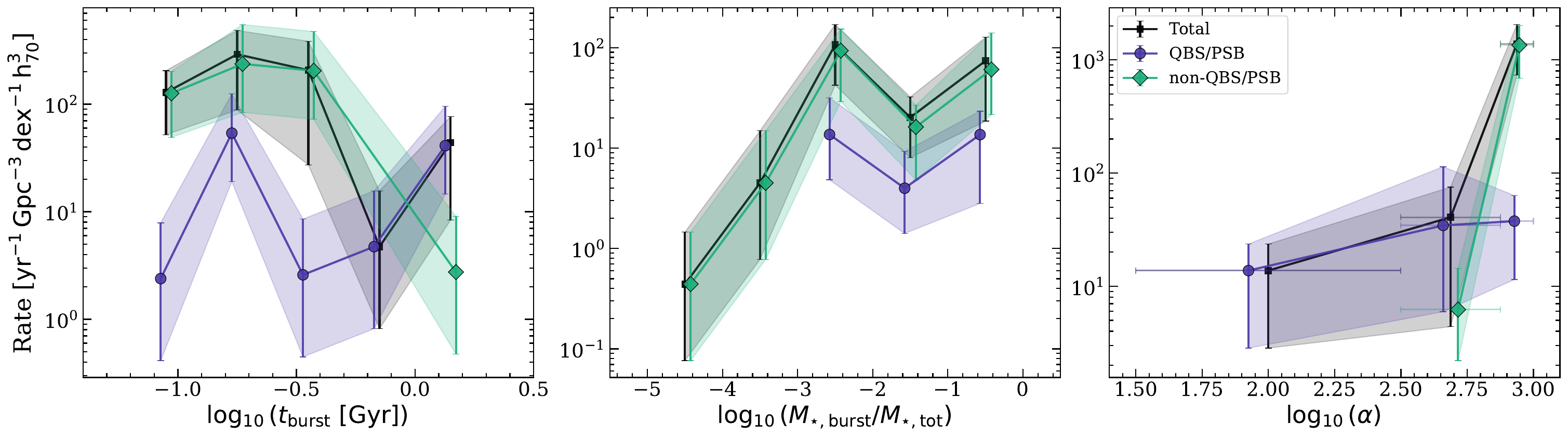}
    \caption{Optical TDE rates compared to the burst age (left panel), the burst mass fraction (middle panel), and $\alpha$, the falling slope of the burst SFH (right panel). The black squares and gray band show the total rate. The purple circles and green diamonds show the rates for TDEs in quiescent Balmer-strong or post-starburst (QBS/PSB) hosts and in all other (non-QBS/PSB) hosts, respectively. The left and right panels include only the 18 hosts with burst mass fractions above 0.1\% (Section~\ref{subsec:burst_props}), while the middle panel includes all 20 hosts. In the right panel, the horizontal bars show the widths of the non-uniform $\alpha$ bins. Subset points are offset slightly from the bin centers for clarity.}
    \label{fig:LF_host_props}
\end{figure*}

In Table~\ref{tab:tde_enhancement}, we quote the full-sample values as our primary result and present the TDE 2023clx-excluded values as a robustness check, since this event carries over a quarter of the total sample weight and is not hosted by a QBS or PSB galaxy. The qualitative conclusion is unchanged in either case: PSB galaxies, and the combined PSB+QBS population, are overrepresented in the volumetric TDE rate at fixed stellar mass and redshift, with the lower bounds of their uncertainties well above $\Gamma = 1$. The QBS-only enhancement, based on three hosts, is of lower significance. These results reinforce the known overrepresentation of QBS and PSB galaxies as TDE hosts \citep[e.g.,][]{Arcavi2014, French2016, French2018, LawSmith2017, Graur2018a}. Using the calibrated TDE weights, we estimate more modest values of $\Gamma$ than many previous works, although they are generally consistent within uncertainties. The slightly lower $\Gamma$ values are likely due to our use of a matched control galaxy sample and weighted TDE frequency.

\subsection{A Recent Burst of Star Formation} \label{subsec:burst_props}

Recent work has investigated the empirical delay time distribution of TDEs to test various theoretical models for TDE rate enhancements in QBS and PSB galaxies \citep{Shepherd2026}. Following a similar procedure (detailed in Section \ref{subsec:host_properties}), we compare the volumetric TDE rates to various properties of any recent burst of star formation. These results are shown in Figure \ref{fig:LF_host_props}, with the time since burst ($t_\textrm{burst}$) shown in the left panel, the burst mass fraction in the middle panel, and $\alpha$, the declining SFH slope of the recent burst component, in the right panel. An important caveat is that galaxies with low estimated burst mass fractions may not have experienced a burst; the recent SFH component likely traces the most recent epoch of star formation. We therefore restrict the following analyses of $t_\textrm{burst}$ and $\alpha$ to host galaxies with burst mass fractions above 0.1\%. In our standard sample, only the host galaxies of TDE 2019dsg and TDE 2022hvp have burst mass fractions below 0.1\%. Each has a large $t_\textrm{burst}$ and low $\alpha$, consistent with a low level of ongoing star formation.

The TDE rate as a function of $t_\textrm{burst}$ is relatively flat, with a slight increase in the volumetric rate at smaller burst ages. An increase in the TDE rate towards lower burst ages broadly aligns with theoretical models predicting enhanced TDE rates within several hundred Myr following a starburst \citep[e.g.,][]{Stone2018, Bortolas2022, Mockler2023, Melchor2024, Wang2024a, Teboul2025}. For QBS and PSB galaxies, all with burst mass fractions $\gtrsim$0.1\%, the rate nominally increases towards larger $t_\textrm{burst}$, reminiscent of the rate enhancements presented in \citet{Shepherd2026}. If we split the sample at a burst mass fraction of 1\% \citep[cf.][]{Shepherd2026}, there is no trend with $t_\textrm{burst}$. Taken together, these trends suggest that multiple factors increase the TDE rate in galaxies with recent bursts of star formation \citep{Shepherd2026}.

The volumetric rate of TDEs is lowest in galaxies that have not recently hosted a burst of star formation. The rate increases by more than two orders of magnitude between a burst mass fraction of $10^{-5}$ and $10^{-2}$. However, above a burst mass fraction of $10^{-3}$, the TDE rate is consistent with a constant rate. This possibly indicates a threshold above which TDE production is more efficient, in broad agreement with the majority of TDE hosts having similarly high burst mass fractions \citep{Shepherd2026}. The total TDE rate is dominated by galaxies with very steeply falling SFHs, indicated by a high $\alpha$, although at most a weak trend exists for QBS/PSB galaxies. This suggests that optically-selected TDEs may prefer rapidly-declining bursts and that dynamical channels for increasing the TDE rate \citep[e.g.,][]{Stone2018, Bortolas2022, Mockler2023, Melchor2024, Teboul2025} are most efficient shortly after a burst quenches.

Finally, the selection of TDE candidates in the IR, many of which lack optical counterparts \citep{Masterson2024}, demonstrates the unsurprising fact that optical selection alone is an incomplete picture of the TDE population. Heavily dust-obscured TDEs will be missed by optical surveys. We examined the rate of TDEs as a function of the fitted A$_V$ in our \texttt{Bagpipes} models and found no correlation between the rate and the amount of dust in the host galaxy (Figure~\ref{fig:LF_vs_dust}, Appendix~\ref{app:A_V}). Although this total A$_V$ is not the line-of-sight dust relevant for obscuring a TDE flare, the similar rate in dusty galaxies suggests that the optical TDE rates are not substantially biased by a modest amount of host-galaxy dust (A$_V$ $\lesssim$ 2 mag). Multi-wavelength selection increases TDE completeness but also probably reduces purity since the cleanest separation between TDEs and other physically distinct accretion flares is based on optical spectroscopic classification.

\subsection{Bulge S\'{e}rsic Index} \label{subsec:host_sersic}

\begin{figure}
    \centering
    \includegraphics[width=\columnwidth]{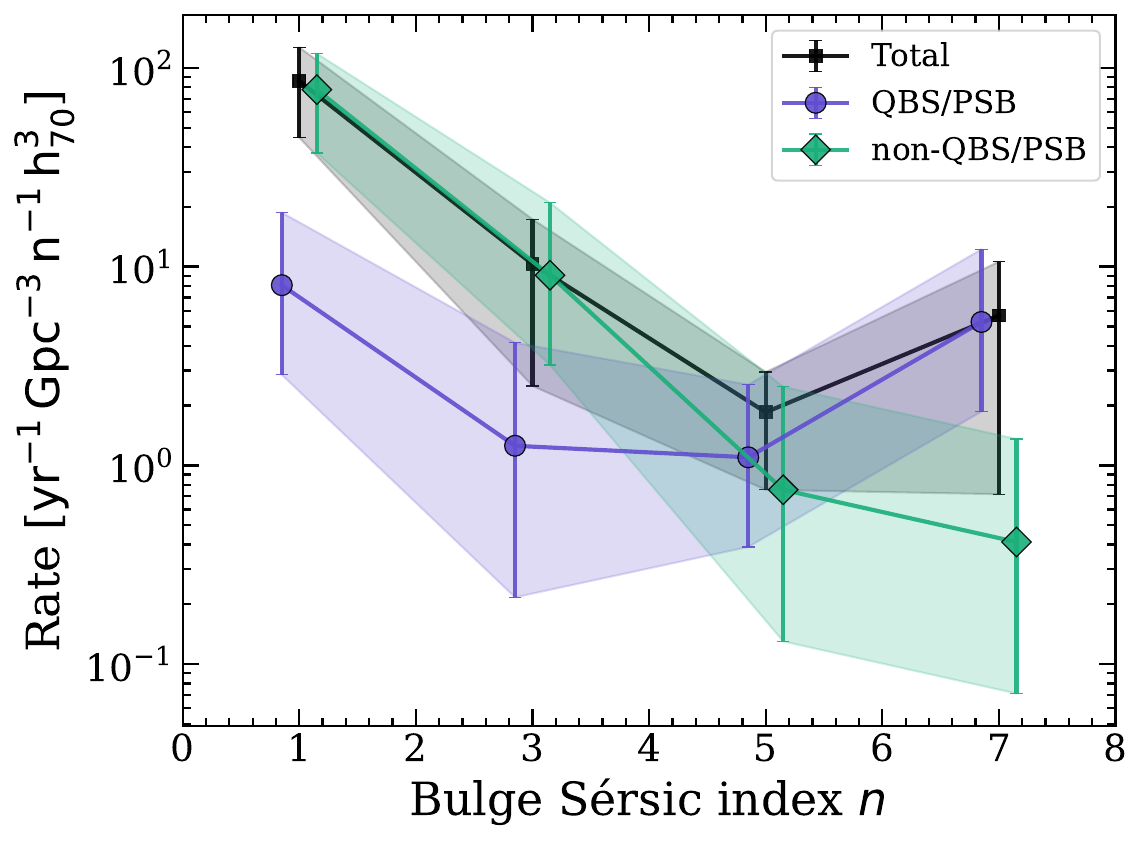}
    \caption{Optical TDE rates as a function of the S\'{e}rsic index $n$ of the host-galaxy bulge, which traces the central concentration of the stellar light profile. The black squares and gray band show the total rate. The purple circles and green diamonds show the rates for TDEs in quiescent Balmer-strong or post-starburst (QBS/PSB) hosts and in all other (non-QBS/PSB) hosts, respectively. Subset points are offset slightly from the bin centers for clarity.}
    \label{fig:LF_vs_sersic}
\end{figure}

Previous studies have shown that TDE host galaxies tend to be centrally concentrated \citep[e.g.,][]{LawSmith2017, Graur2018a, Hammerstein2021, Newsome2025, Chang2026}. In Figure \ref{fig:LF_vs_sersic}, we examine the volumetric TDE rate as a function of the bulge S\'ersic index $n$. The TDE rate is dominated by galaxies with small $n$ and continues to decrease with increasing $n$ before possibly plateauing for $n \gtrsim 5$. A S\'ersic index of $n \simeq 2$ is often used to separate pseudobulges ($n \lesssim 2$) and classical bulges ($n \gtrsim 2$), which may form through secular processes and mergers, respectively \citep[e.g.,][]{Fisher2008, Gadotti2009}. This matches the known prevalence of TDEs in low-mass host galaxies where major mergers are less common \citep[e.g.,][]{French2020b, vanVelzen2021, Hammerstein2021}, although some TDE host galaxies do exhibit morphological signatures consistent with recent mergers \citep[e.g.,][]{Prieto2016}. The high TDE rate in galaxies that likely host pseudobulges is also consistent with \citet{Chang2026}, who find that the majority of TDE hosts lack signs of morphological disturbances and are more likely than matched control galaxies to show bar- or ring-like structures, suggesting that their recent evolution has been dominated by bar-driven secular processes rather than mergers.

The plateau in the TDE rate for galaxies with large $n$ is primarily associated with QBS and PSB hosts. In fact, all galaxies with $n > 5$ are QBS or PSB after considering the host galaxy of TDE 2019ahk (ASASSN-19bt). TDE 2019ahk occurred in a galaxy with H$\alpha$ EW $=27.97 \pm 0.80$ \AA\ and Lick H$\delta_A = 6.08 \pm 0.92$ \AA, which places it near the population of shocked post-starburst galaxies \citep[SPOGs;][]{Alatalo2016, Holoien2019c}. Although \citet{Holoien2019c} could not robustly identify the host galaxy of TDE 2019ahk as a member of the SPOG population, the high burst mass fraction at a $\sim$Gyr-old age (see Table \ref{tab:tde_sample_hostprops}) supports its classification as a SPOG. The plateau in the total volumetric TDE rate at large $n$ is therefore plausibly associated with the overrepresentation of QBS and PSB galaxies.

\subsection{Supermassive Black Hole Mass} \label{subsec:smbh_mass_func}

\begin{figure}
    \centering
    \includegraphics[width=\columnwidth]{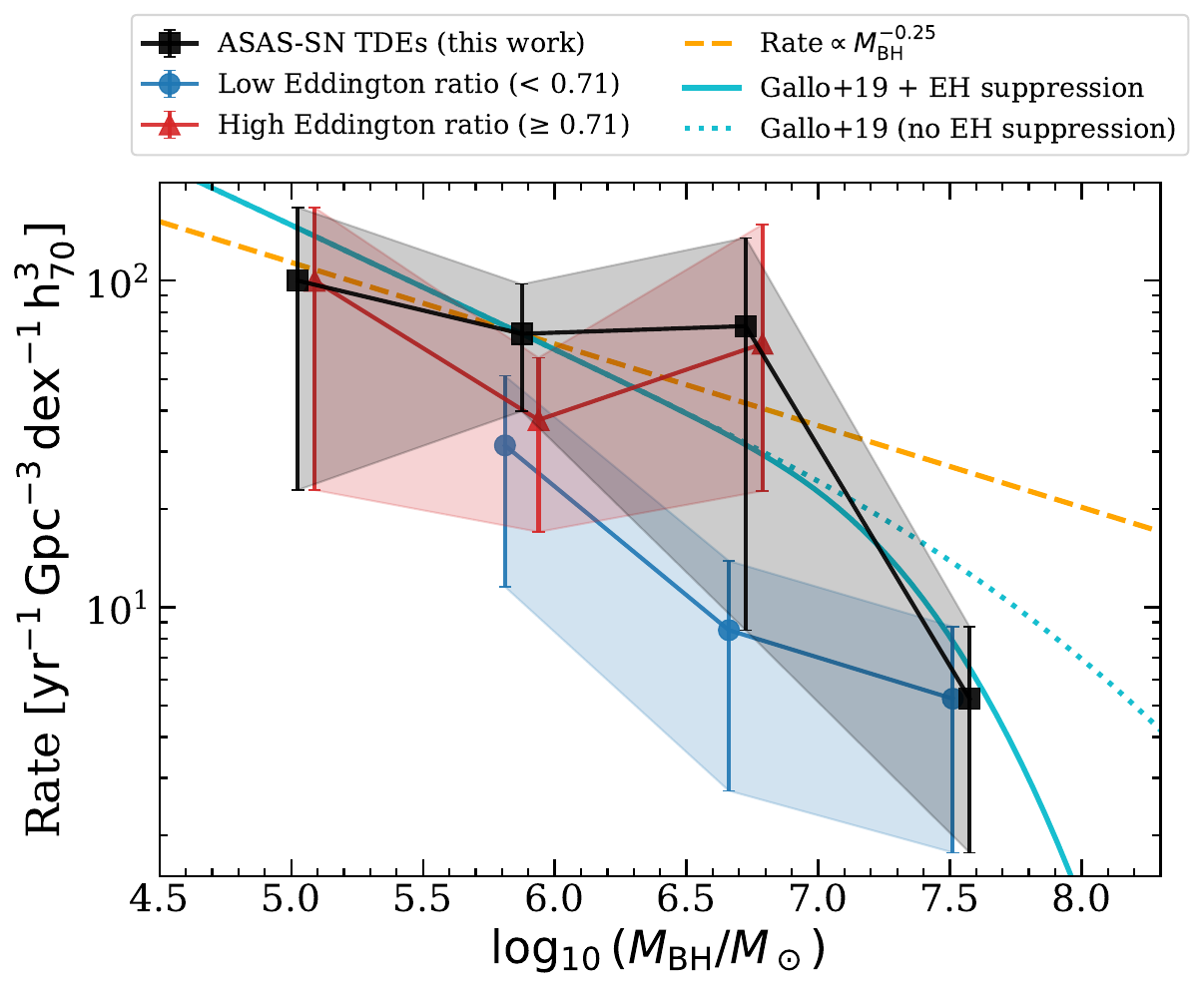}
    \caption{Optical TDE rates as a function of the SMBH mass. The rates are colored by the UV/optical Eddington ratio of the TDE, split at the median Eddington ratio of the sample. Subset points are offset slightly from the bin centers for clarity. We recover the $M_\mathrm{BH}^{-0.25}$ scaling (orange dashed line) measured by \citet{Yao2023}. We additionally show two expectations for the TDE SMBH mass function from \citet{Gallo2019} with (solid) and without (dotted) event horizon suppression. All three comparison curves are normalized to our total measured rate in the bin centered at $\log_{10}(M_\textrm{BH}/M_\odot) = 5.875$.}
    \label{fig:LF_vs_Mbh}
\end{figure}

The SMBH mass on which a TDE occurs is fundamental in setting key observational parameters, including the timescale \citep[e.g.,][]{Wang2004, Stone2016, Kochanek2016b} and may regulate the observed bolometric luminosity through Eddington-limited accretion and reprocessing \citep[e.g.,][]{Wevers2017, Wevers2019b}. Thus, it is an important parameter for rate comparisons \citep{vanVelzen2018, Yao2023}. TDEs have long been predicted to occur at higher rates on lower-mass SMBHs \citep[e.g.,][]{Magorrian1999, Wang2004, Stone2016, Kochanek2016b} and the precipitous drop in the rates at the highest SMBH masses is well-known \citep[e.g.,][]{Hills1975, vanVelzen2018}. TDEs on IMBHs are expected \citep[e.g.,][]{MacLeod2016b, Chang2025}, although the rate predictions vary. Because the tidal disruption radius approaches the stellar radius below $\sim$$10^{5}$ M$_\odot$, there may be qualitative changes in TDE phenomenology in the IMBH regime. Figure \ref{fig:LF_vs_Mbh} shows the volumetric rate of optical TDEs compared to their SMBH mass. As discussed in Section \ref{subsec:smbh_masses}, we favor masses estimated through late-time TDE observations.

Like previous measurements of the TDE SMBH mass function, we find that the TDE rate is relatively flat below $M_\mathrm{BH} \sim 10^{7}\,M_\odot$ \citep{vanVelzen2018, Yao2023}. At low SMBH masses, our mass function is consistent with the rate $\propto M_\mathrm{BH}^{-0.25}$ dependence found by \citet{Yao2023}, suggestive of a flat SMBH mass function. Similarly, we find that TDE rates drop for $M_\mathrm{BH} \gtrsim 10^{7.5}\,M_\odot$, broadly consistent with the expected suppression near and above the Hills mass for low-mass stars. The observed Eddington ratios are highest around less massive SMBHs and drop for higher-mass SMBHs, consistent with the expected fallback rates over this mass range \citep[e.g.,][]{Ulmer1999, Strubbe2009, Kochanek2016b, Stone2016}. The high Eddington-ratio TDEs dominate the total rate, confirming that local TDEs are excellent probes of super-Eddington accretion.

ANTs and ENTs are promising transient probes of high-mass SMBHs \citep[e.g.,][]{Subrayan2023, Wiseman2025, Hinkle2025}. An extrapolation of the high-mass end of the TDE SMBH mass function, following the \citet{Gallo2019} functional form with event horizon suppression, is again broadly consistent with existing rate estimates for luminous ANTs and ENTs. The estimated ANT rate of $\gtrsim 3\times10^{-2}$ yr$^{-1}$ Gpc$^{-3}$ \citep[][]{Wiseman2025} agrees with our TDE SMBH mass function at $\log_{10}(M_{\rm BH}/M_{\odot}) \approx 8.6$. The general agreement in rates supports their interpretation as TDEs in AGNs and/or massive star TDEs \citep[][]{Subrayan2023, Wiseman2025, Hinkle2025, Graham2026}.

\section{Summary} \label{sec:summary}
 
We present the volumetric rate of optical TDEs as measured from the ASAS-SN survey. Our standard sample comprises 21 TDEs within a redshift of $z \le 0.12$, spanning $g$-band absolute magnitudes from $-17$ to $-22$~mag. The ASAS-SN completeness is calibrated and accounted for when measuring the TDE rate. We additionally present the rates of two interesting subpopulations: off-nuclear and recurrent TDEs. We also examine the dependence of the optical LFs on spectroscopic type, X-ray luminosity, UV/optical blackbody properties, and host-galaxy properties. Our conclusions are summarized as follows:

\begin{itemize}

\item We measure a total local volumetric rate of optical TDEs of $\log_{10}(R_\mathrm{tot}\,[\Runit]) = 2.32^{+0.15}_{-0.23}$ ($R_\mathrm{tot} = 210 \pm 85$~$\Runit$). The ASAS-SN $g$-band LF agrees in shape with previous estimates \citep[e.g.,][]{vanVelzen2018, Lin2022, Charalampopoulos2023, Yao2023}, although its normalization is lower than that of \citet{Yao2023} by a factor of $\sim$2 at $L_g \gtrsim 10^{43}$ erg s$^{-1}$.

\item Since the off-nuclear TDE 2024tvd (ZTF22aaigqsr) is in our standard sample, we estimate the local volumetric rate of off-nuclear TDEs to be $\log_{10}(R_{\textrm{off-nuclear}}\,[\Runit]) = 0.39^{+0.52}_{-0.76}$. This rate is $1.2^{+3.4}_{-1.0}$\% of the total rate, consistent with existing coarse estimates \citep{Stein2026}.

\item There are 4 recurrent TDEs in our standard sample: TDE 2019azh (ASASSN-19dj), TDE 2018fyk (ASASSN-18ul), TDE 2022dbl (ASASSN-22ci), and TDE 2024pvu (ZTF18abxnvoz), which are plausibly also repeating partial TDEs \citep[e.g.,][]{Yao2026a, Hinkle2026}. The rTDE rate is $\log_{10}(R_{\rm recurrent}\,[\Runit]) = 1.26^{+0.23}_{-0.62}$ ($9^{+8}_{-6}$\% of the total rate). Including the well-studied rTDE ASASSN-14ko, the rate is $\log_{10}(R_{\rm recurrent}\,[\Runit]) = 1.45^{+0.20}_{-0.35}$, with a fractional rate of $13^{+10}_{-7}$\%. Our central value of $\approx$9--13\% is lower than the previous estimates ($25\mbox{--}60$\%; \citealt{Yao2026a}). If each flare of the rTDEs is counted as an independent TDE in the same host galaxy, the inferred rTDE rate rises to $\log_{10}(R_{\rm recurrent,\,flares}\,[\Runit]) = 1.91^{+0.25}_{-0.72}$ ($30^{+19}_{-23}\%$ of all TDE flares).

\item The volumetric rate of optically-selected TDEs is dominated by galaxies with significant ($>$0.1\%) burst-mass fractions and steeply falling SFHs.

\item The volumetric rate of TDEs predominantly follows the number density of galaxies in a redshift- and mass-matched control sample. However, known rate enhancements are recovered for PSB galaxies ($\Gamma = 15^{+16}_{-11}$) and, at lower significance, for QBS galaxies ($\Gamma = 5^{+5}_{-4}$).

\item TDE-H+He events dominate the total TDE rate, with TDE-H and TDE-He events comprising 3.4\% of the overall optical TDE rate. Luminous ($M < -20$ mag) TDE-Featureless events are $<$0.5\% of local TDEs. Low-luminosity TDE-Featureless events are not well-constrained, but may represent $\sim$10\% of the total TDE rate.

\item The $g$-band LFs are statistically indistinguishable for X-ray bright and X-ray faint events, supporting unification models where the viewing angle sets the multi-wavelength appearance of the TDE \citep[e.g.,][]{Dai2018a, Thomsen2022}.

\item We recover a known $M_\mathrm{BH}^{-0.25}$ dependence of the volumetric rates, suggestive of a flat SMBH mass distribution \citep[$dn_\mathrm{BH}/d\log M_\mathrm{BH} \approx \mathrm{constant}$;][]{Yao2023}.

\item Extrapolations of the optical volumetric TDE rate to higher luminosities and SMBH masses are broadly consistent with the estimated rates of nuclear flares like luminous ANTs \citep[e.g.,][]{Wiseman2025} and ENTs \citep[e.g.,][]{Hinkle2025}. This tentatively supports claims that they are powered by TDEs in AGNs and/or TDEs of massive stars \citep[e.g.,][]{Hinkle2024b, Wiseman2025, Hinkle2025, Graham2026, Quintin2026}.

\end{itemize}
 
A robustly estimated volumetric TDE rate provides important information on the physics of TDEs themselves (see Section \ref{sec:gband_LF} and Section \ref{sec:BB_rates}) and insights into the relationship between host-galaxy properties and TDE production (see Section \ref{sec:hostgal_rates}). Existing surveys provide well-constrained local volumetric TDE rates, a key input for estimating the expected yield \citep[e.g.,][]{Bricman2020, Shvartzvald2024, Karmen2026} from the new generation of surveys such as the Legacy Survey of Space and Time on the Vera C. Rubin Observatory \citep[LSST;][]{Ivezic2019}, the Roman Space Telescope's High-Latitude Time-Domain Survey \citep{Rose2021}, and the Ultraviolet Transient Astronomy Satellite \citep[ULTRASAT;][]{Shvartzvald2024}. These surveys will constrain the redshift evolution of TDE rates and eventually probe the SMBH mass function at higher redshifts \citep[e.g.,][]{Karmen2026}. Sufficiently large TDE samples may also provide a path towards constraining the spin distribution of quiescent SMBHs \citep[e.g.,][]{Kesden2012, Huang2024a}, although there are degeneracies with massive star TDEs. In any case, local TDE samples will continue to complement these large higher-redshift samples by providing the most detailed view of the physics of TDEs and their host galaxies.

\section*{Acknowledgments}
 
J.T.H. acknowledges support from NASA through the NASA Hubble Fellowship grant HST-HF2-51577.001-A, awarded by STScI. STScI is operated by the Association of Universities for Research in Astronomy, Incorporated, under NASA contract NAS5-26555.
The Shappee group at the University of Hawai`i is supported with funds from NSF grant AST-2407205.
ASAS-SN is funded by Gordon and Betty Moore Foundation grants GBMF5490 and GBMF10501 and the Alfred P. Sloan Foundation grant G-202114192. 
C.S.K. and K.Z.S. are supported by NSF grants AST-2307385 and AST-2407206. 
M.S. and K.D.F. acknowledge support from NSF grant AST 22-06164. M.S. thanks the LSST-DA Data Science Fellowship Program, which is funded by LSST-DA, the Brinson Foundation, the WoodNext Foundation, and the Research Corporation for Science Advancement Foundation; her participation in the program has benefited this work. 
J.F.B. is supported by NSF grant PHY-2609870 and was supported in part by the World Research Hub
(WRH) Program of Institute of Science Tokyo. 
S.D. is supported by the National Natural Science Foundation of China (Grant No.\ 12133005). 
W.B.H. acknowledges support from the NSF Graduate Research Fellowship Program under Grant No. 2236415.
Parts of this research were supported by the Australian Research Council Discovery Early Career Researcher Award (DECRA) through project number DE230101069.
D.O.J. acknowledges support from NSF grants AST-2407632, AST-2429450, AST-2510993, and AST-2607081, NASA grants 80NSSC24M0023 and 80NSSC24K0353, and JWST-GO-05324.031, awarded by the Space Telescope Science Institute (STScI), which is operated by the Association of Universities for Research in Astronomy, Inc., for NASA, under contract NAS5-26555.  This work is also funded in part by the Gordon and Betty Moore Foundation through Grant GBMF13900 to D.O.J.

Funding for SDSS-III has been provided by the Alfred P. Sloan Foundation, the Participating Institutions, the National Science Foundation, and the U.S. Department of Energy Office of Science. The SDSS-III web site is http://www.sdss3.org/.
 
SDSS-III is managed by the Astrophysical Research Consortium for the Participating Institutions of the SDSS-III Collaboration including the University of Arizona, the Brazilian Participation Group, Brookhaven National Laboratory, Carnegie Mellon University, University of Florida, the French Participation Group, the German Participation Group, Harvard University, the Instituto de Astrofisica de Canarias, the Michigan State/Notre Dame/JINA Participation Group, Johns Hopkins University, Lawrence Berkeley National Laboratory, Max Planck Institute for Astrophysics, Max Planck Institute for Extraterrestrial Physics, New Mexico State University, New York University, Ohio State University, Pennsylvania State University, University of Portsmouth, Princeton University, the Spanish Participation Group, University of Tokyo, University of Utah, Vanderbilt University, University of Virginia, University of Washington, and Yale University.

Some of the data presented herein were obtained at Keck Observatory, which is a private 501(c)3 non-profit organization operated as a scientific partnership among the California Institute of Technology, the University of California, and the National Aeronautics and Space Administration. The Observatory was made possible by the generous financial support of the W. M. Keck Foundation. This research has made use of the Keck Observatory Archive (KOA), which is operated by the W. M. Keck Observatory and the NASA Exoplanet Science Institute (NExScI), under contract with the National Aeronautics and Space Administration.

This publication makes use of data products from the Two Micron All Sky Survey, which is a joint project of the University of Massachusetts and the Infrared Processing and Analysis Center/California Institute of Technology, funded by the National Aeronautics and Space Administration and the National Science Foundation.

We acknowledge the use of public data from the Swift data archive.

The Legacy Surveys consist of three individual and complementary projects: the Dark Energy Camera Legacy Survey (DECaLS; Proposal ID \#2014B-0404; PIs: David Schlegel and Arjun Dey), the Beijing-Arizona Sky Survey (BASS; NOAO Prop. ID \#2015A-0801; PIs: Zhou Xu and Xiaohui Fan), and the Mayall z-band Legacy Survey (MzLS; Prop. ID \#2016A-0453; PI: Arjun Dey). DECaLS, BASS and MzLS together include data obtained, respectively, at the Blanco telescope, Cerro Tololo Inter-American Observatory, NSF’s NOIRLab; the Bok telescope, Steward Observatory, University of Arizona; and the Mayall telescope, Kitt Peak National Observatory, NOIRLab. Pipeline processing and analyses of the data were supported by NOIRLab and the Lawrence Berkeley National Laboratory (LBNL). The Legacy Surveys project is honored to be permitted to conduct astronomical research on I’oligam Du’ag (Kitt Peak), a mountain with particular significance to the Tohono O’odham Nation.

NOIRLab is operated by the Association of Universities for Research in Astronomy (AURA) under a cooperative agreement with the National Science Foundation. LBNL is managed by the Regents of the University of California under contract to the U.S. Department of Energy.

This project used data obtained with the Dark Energy Camera (DECam), which was constructed by the Dark Energy Survey (DES) collaboration. Funding for the DES Projects has been provided by the U.S. Department of Energy, the U.S. National Science Foundation, the Ministry of Science and Education of Spain, the Science and Technology Facilities Council of the United Kingdom, the Higher Education Funding Council for England, the National Center for Supercomputing Applications at the University of Illinois at Urbana-Champaign, the Kavli Institute of Cosmological Physics at the University of Chicago, Center for Cosmology and Astro-Particle Physics at the Ohio State University, the Mitchell Institute for Fundamental Physics and Astronomy at Texas A\&M University, Financiadora de Estudos e Projetos, Fundacao Carlos Chagas Filho de Amparo, Financiadora de Estudos e Projetos, Fundacao Carlos Chagas Filho de Amparo a Pesquisa do Estado do Rio de Janeiro, Conselho Nacional de Desenvolvimento Cientifico e Tecnologico and the Ministerio da Ciencia, Tecnologia e Inovacao, the Deutsche Forschungsgemeinschaft and the Collaborating Institutions in the Dark Energy Survey. The Collaborating Institutions are Argonne National Laboratory, the University of California at Santa Cruz, the University of Cambridge, Centro de Investigaciones Energeticas, Medioambientales y Tecnologicas-Madrid, the University of Chicago, University College London, the DES-Brazil Consortium, the University of Edinburgh, the Eidgenossische Technische Hochschule (ETH) Zurich, Fermi National Accelerator Laboratory, the University of Illinois at Urbana-Champaign, the Institut de Ciencies de l’Espai (IEEC/CSIC), the Institut de Fisica d’Altes Energies, Lawrence Berkeley National Laboratory, the Ludwig Maximilians Universitat Munchen and the associated Excellence Cluster Universe, the University of Michigan, NSF’s NOIRLab, the University of Nottingham, the Ohio State University, the University of Pennsylvania, the University of Portsmouth, SLAC National Accelerator Laboratory, Stanford University, the University of Sussex, and Texas A\&M University.

BASS is a key project of the Telescope Access Program (TAP), which has been funded by the National Astronomical Observatories of China, the Chinese Academy of Sciences (the Strategic Priority Research Program “The Emergence of Cosmological Structures” Grant \# XDB09000000), and the Special Fund for Astronomy from the Ministry of Finance. The BASS is also supported by the External Cooperation Program of Chinese Academy of Sciences (Grant \# 114A11KYSB20160057), and Chinese National Natural Science Foundation (Grant \# 12120101003, \# 11433005).

The Legacy Survey team makes use of data products from the Near-Earth Object Wide-field Infrared Survey Explorer (NEOWISE), which is a project of the Jet Propulsion Laboratory/California Institute of Technology. NEOWISE is funded by the National Aeronautics and Space Administration.

The Legacy Surveys imaging of the DESI footprint is supported by the Director, Office of Science, Office of High Energy Physics of the U.S. Department of Energy under Contract No. DE-AC02-05CH11231, by the National Energy Research Scientific Computing Center, a DOE Office of Science User Facility under the same contract; and by the U.S. National Science Foundation, Division of Astronomical Sciences under Contract No. AST-0950945 to NOAO.

The authors wish to recognize and acknowledge the very significant cultural role and reverence that the summits of Maunakea and Haleakal\=a have always had within the Native Hawaiian community. We are most fortunate to have the opportunity to conduct observations from this mountain.

\bibliography{references}
\bibliographystyle{mnras}

%-----------------------------------------------------------------------------%
\begin{appendices}
%-----------------------------------------------------------------------------%
\appendix
% 1. Reset the figure/table counter whenever a new section starts
% 2. Change the numbering format to SectionLetter + Number (e.g., A1)
\counterwithin{figure}{section}
\counterwithin{table}{section}
\renewcommand{\thefigure}{\thesection\arabic{figure}}
\renewcommand{\thetable}{\thesection\arabic{table}}
\renewcommand{\theHequation}{\thesection\arabic{equation}}
\renewcommand{\theHfigure}{\thesection\arabic{figure}}
\renewcommand{\theHtable}{\thesection\arabic{table}}

\section{Rates vs \texorpdfstring{$A_V$}{A\_V}} \label{app:A_V}

\begin{figure}[h]
    \centering
    \includegraphics[width=\columnwidth]{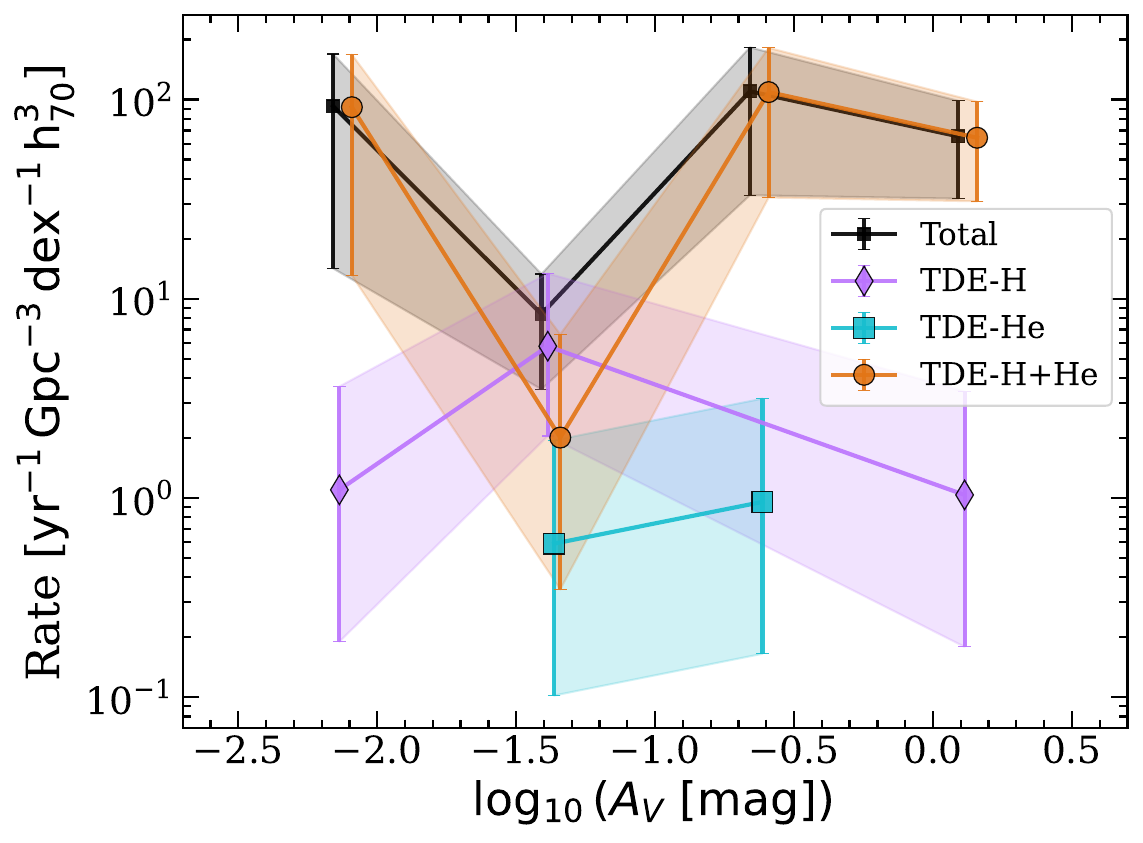}
    \caption{Optical TDE rates as a function of the host-galaxy dust attenuation $A_V$ from our \texttt{Bagpipes} SED modeling (Section~\ref{subsec:host_properties}). The black squares show the total rate. Purple diamonds, cyan squares, and orange circles show TDE-H, TDE-He, and TDE-H+He events, respectively. The rate is consistent with being flat as a function of $A_V$. Subset points are offset slightly from the bin centers for clarity.}
    \label{fig:LF_vs_dust}
\end{figure}

Figure~\ref{fig:LF_vs_dust} shows the volumetric TDE rate as a function of the host-galaxy dust attenuation $A_V$ from our \texttt{Bagpipes} modeling (Section~\ref{subsec:host_properties}). The rates are computed exactly as in Section~\ref{sec:rates}. Each TDE enters its $A_V$ bin with the statistical weight $w_i = (F_{1,i}\,F_{2,i}\,F_{3,i})^{-1}$, and uncertainties come from the same joint Poisson bootstrap used throughout. Of the 21 TDEs in our standard sample, 20 have host-galaxy SED fits; TDE~2024tvd is excluded as an off-nuclear event. The fitted attenuations span $A_V \approx 0.005$--$2$~mag. The rate is consistent with being flat across the full range, both in total and for the individual spectroscopic types.
 
\section{Source of Data} \label{app:sources}

The sources of the data used in this work are not homogeneous. Table~\ref{tab:extra_host_info} lists, for every TDE in the sample, the first publication (or TNS classification report) identifying it as a TDE and the sources of its host-galaxy properties, X-ray luminosity, and SMBH mass. For hosts previously analyzed in \cite{Shepherd2026}, the original $M_{\star,\rm tot}$ and host spectrum sources are given in their Table~D1, and the H$\alpha$ EW and Lick H$\delta_A$ sources in their Section~3.2, unless otherwise noted in Table~\ref{tab:extra_host_info}.

\begin{table*}
    \centering
    \renewcommand{\arraystretch}{1.4}
    \setlength{\tabcolsep}{5pt}
    \caption{Sources of the data used in this work.}
    \label{tab:extra_host_info}
    \resizebox{\textwidth}{!}{%
    \begin{tabular}{l l c c c c c c c}
    \toprule
    {IAU Name} & {Discovery Name} & {Disc./Class.} & {$M_{\star,\rm tot}$} & {H$\alpha$ EW} & {Lick H$\delta_A$} & {Host spectrum} & {$L_{\rm X}$} & {$M_{\rm BH}$} \\
    \midrule
    \nodata & ASASSN-14ae & 1 & 2 & 2 & 2 & 2 & 1 & 3 \\
    \nodata & ASASSN-14li & 4 & 2 & 2 & 2 & 2 & 4 & 3 \\
    \nodata & ASASSN-15oi & 5 & 2 & 2 & 2 & 2 & 6 & 3 \\
    2018dyb & ASASSN-18pg & 7 & 2 & 2 & 2 & 2 & 8 & 3 \\
    2018fyk & ASASSN-18ul & 9 & 2 & 2 & 10 & 2 & 9 & 3 \\
    2018hyz & ASASSN-18zj & 11 & 2 & 2 & 2 & 2 & 12 & 3 \\
    2018zr & PS18kh & 13 & 14 & This work & This work & This work$^{a}$ & 14 & 3 \\
    2019ahk & ASASSN-19bt & 15 & 2 & 2 & 15 & 2 & 15 & 3 \\
    2019azh & ASASSN-19dj & 16 & 2 & 2 & 2 & 2 & 17 & 3 \\
    2019dsg & ZTF19aapreis & 16 & 2 & 2 & This work$^{a}$ & 2 & 18 & 3 \\
    2019qiz & ZTF19abzrhgq & 16 & 2 & 2 & 2 & 2 & 19$^{b}$ & 3 \\
    2019vcb & ZTF19acspeuw & 20 & 20 & This work & This work & This work$^{a}$ & 21$^{b}$ & 22$^{c}$ \\
    2020neh & ZTF20abgwfek & 23 & 24 & 24 & 24 & 24 & 23 & 3 \\
    2020nov & ZTF20abisysx & 25 & 2 & 2 & 2 & 2 & 25 & 25 \\
    2020vwl & Gaia20etp & 22 & 2 & 2 & 2 & 2 & This work$^{d}$ & 3 \\
    2022bdw & ATLAS22dth & 26 & 2 & 2 & 2 & 2 & This work$^{d}$ & 27 \\
    2022dbl$^{e}$ & ASASSN-22ci & 28 & 2 & 2 & 2 & 2 & 10 & 27 \\
    2022dsb & ASASSN-22cs & 29 & 29 & This work & This work & This work$^{f}$ & 29 & 27 \\
    2022hvp & ZTF22aagyuao & 30 & 30 & This work & This work & This work$^{g}$ & This work$^{d}$ & 27 \\
    2022lri & ATLAS22pnz & 31 & 31 & This work & This work & This work$^{a}$ & 31 & 31$^{h}$ \\
    2023clx & ASASSN-23bd & 32 & 2 & 2 & 2 & 2 & 33 & 27 \\
    2023mhs & ZTF23aaqdjhi & 34 & 2 & 2 & 2 & 2 & This work$^{d}$ & 2$^{c}$ \\
    2024as & ATLAS24acr & 35 & This work$^{i}$ & This work & This work & This work$^{g}$ & This work$^{d}$ & This work$^{c}$ \\
    2024pvu & ZTF18abxnvoz & 36 & 37 & This work & This work & This work$^{g}$ & This work$^{d}$ & 37 \\
    2024tvd$^{j}$ & ZTF22aaigqsr & 38 & \nodata & \nodata & \nodata & \nodata & 38 & 39 \\
    \bottomrule
    \end{tabular}%
    }
    
    \vspace{4pt}
    \begin{minipage}{\textwidth}
    \raggedright
    References: (1)~\cite{Holoien2014b}; (2)~\cite{Shepherd2026}; (3)~\cite{Mummery2024a}; (4)~\cite{Holoien2016a}; (5)~\cite{Holoien2016c}; (6)~\cite{Holoien2018}; (7)~\cite{Leloudas2019}; (8)~\cite{Holoien2020}; (9)~\cite{Wevers2019b}; (10)~\cite{Hinkle2024c}; (11)~\cite{Short2020}; (12)~\cite{Hung2020}; 
    (13)~\cite{vanVelzen2019}; (14)~\cite{Holoien2019b}; (15)~\cite{Holoien2019c}; (16)~\cite{vanVelzen2021}; (17)~\cite{Hinkle2021a}; (18)~\cite{Cannizzaro2021}; (19)~\cite{Nicholl2020}; (20)~\cite{Hammerstein2023}; (21)~\cite{Quintin2023}; (22)~\cite{Yao2023}; (23)~\cite{Angus2022}; (24)~SDSS {\tt galSpec} \citep{Kauffmann2003a, Brinchmann2004, Tremonti2004}; (25)~\cite{Earl2025}; (26)~\cite{Arcavi2022TNS}; (27)~\cite{Mummery2025}; (28)~\cite{Lin2024}; (29)~\cite{Malyali2024}; (30)~\cite{Yao2025b}; (31)~\cite{Yao2024b}; (32)~\cite{Zhu2023}; (33)~\cite{Hoogendam2024b}; (34)~\cite{Newsome2023TNS}; (35)~\cite{Brivio2024TNS}; (36)~\cite{Jordana-Mitjans2025}; (37)~\cite{Yao2026a}; (38)~\cite{Yao2025a}; (39)~\cite{Patra2025}.\\
    \vspace{4pt}
    $^{a}$ Keck/LRIS spectrum. \\
    $^{b}$ Excluding the quasi-periodic eruption (QPE) emission. \\
    $^{c}$ Estimated from the host stellar mass of the cited source using the $M_{\rm BH}$--$M_{\star}$ relation of \cite{Reines2015}. \\
    $^{d}$ Estimated from Swift/XRT data. \\
    $^{e}$ Listed as AT2018mac in \cite{Shepherd2026}. \\
    $^{f}$ Keck/KCWI spectrum. \\
    $^{g}$ UH88/SNIFS spectrum obtained as part of the Spectroscopic Classification of Astronomical Transients (SCAT) survey \citep{Tucker2022b, Tucker2026}. \\
    $^{h}$ From the $M_{\rm BH}$--$\sigma_\star$ relation. \\
    $^{i}$ As in \cite{Shepherd2026}, to estimate the total stellar mass of the galaxy from the 2MASS \citep{Skrutskie2006} K-band apparent magnitude, we calculate the distance to the galaxy using the \texttt{astropy} \texttt{cosmo.distmod} method. We use the distance and the apparent magnitude to calculate the absolute magnitude, converting to luminosity. Using the mass-to-light ratio in \cite{Just2015}, we calculate the stellar mass of the host galaxy. \\
    $^{j}$ Off-nuclear TDE; its host-galaxy properties are not used in this work. \\
    \textit{Note.} ``This work'' entries for H$\alpha$ EW and Lick H$\delta_A$ were measured with \texttt{Bagpipes} and \texttt{pyLick}, respectively (Section~\ref{subsec:host_properties}).
    \end{minipage}
\end{table*}

\section{Binned Rates} \label{app:binned_rates}

Table~\ref{tab:lf_gband} lists the binned $g$-band LF shown in Figure~\ref{fig:gband_LF}, and Table~\ref{tab:lf_bb} lists the binned peak blackbody-luminosity LF (Figure~\ref{fig:LF_BB_props}, left panel), both for the total sample and by spectroscopic type. Table~\ref{tab:enhancement_all} gives the TDE rate and rate enhancement in every region of the H$\alpha$ EW--Lick H$\delta_A$ plane (Section~\ref{subsec:host_qbs_psb}). All other binned rates shown in this paper, including those for each subset, will be made available in machine-readable form on Zenodo upon publication.

\begin{table*}
\centering
\caption{Binned peak $g$-band LF (Figure~\ref{fig:gband_LF}), given as $\log_{10}(\phi\,[\Runit\,\mathrm{dex^{-1}}])$ with $\phi = dR/d\log_{10}L_g$, for the total standard sample, each spectroscopic type, and the two X-ray classes. Uncertainties are the 16th--84th percentiles of the bootstrap.}
\label{tab:lf_gband}
\renewcommand{\arraystretch}{1.7}
\setlength{\tabcolsep}{4.4pt}
\begin{tabular}{l | c c c c c c c}
\toprule
 & & \multicolumn{6}{c}{$\log_{10}(\phi\,[\Runit\,\mathrm{dex^{-1}}])$} \\
\cmidrule(lr){3-8}
$\log_{10}(L_g/\mathrm{erg\,s^{-1}})$ & $N$ & Total & TDE-H & TDE-He & TDE-H+He & $L_\mathrm{X}/L_\mathrm{BB} < 0.01$ & $L_\mathrm{X}/L_\mathrm{BB} \geq 0.01$ \\
\midrule
42.2--42.6 & 1 & $2.16^{+0.52}_{-0.76}$ & \nodata & \nodata & $2.16^{+0.52}_{-0.76}$ & $2.16^{+0.52}_{-0.76}$ & \nodata \\
42.6--43.0 & 7 & $2.50^{+0.17}_{-0.29}$ & \nodata & \nodata & $2.50^{+0.17}_{-0.29}$ & $2.05^{+0.18}_{-0.32}$ & $2.31^{+0.22}_{-0.62}$ \\
43.0--43.4 & 7 & $1.71^{+0.15}_{-0.23}$ & $0.90^{+0.52}_{-0.76}$ & \nodata & $1.63^{+0.16}_{-0.26}$ & $1.30^{+0.37}_{-0.45}$ & $1.49^{+0.17}_{-0.29}$ \\
43.4--43.8 & 5 & $0.98^{+0.17}_{-0.29}$ & $0.83^{+0.20}_{-0.38}$ & $0.25^{+0.52}_{-0.76}$ & $-0.04^{+0.52}_{-0.76}$ & $0.83^{+0.20}_{-0.38}$ & $0.43^{+0.37}_{-0.45}$ \\
43.8--44.2 & 1 & $0.04^{+0.52}_{-0.76}$ & \nodata & $0.04^{+0.52}_{-0.76}$ & \nodata & $0.04^{+0.52}_{-0.76}$ & \nodata \\
\bottomrule
\end{tabular}

\vspace{4pt}
\begin{minipage}{\textwidth}
\raggedright
\textit{Note.} $N$ is the number of TDEs in the bin. Bins in which a subset has no events are marked with an ellipsis. Subsets with $N \leq 2$ events in a bin use the \citet{Gehrels1986} $1\sigma$ limits scaled to the measured rate (Section~\ref{subsec:vol_rates}).
\end{minipage}
\end{table*}

\begin{table*}
\centering
\caption{Binned peak blackbody-luminosity LF (Figure~\ref{fig:LF_BB_props}, left panel), given as $\log_{10}(\phi\,[\Runit\,\mathrm{dex^{-1}}])$ with $\phi = dR/d\log_{10}L_\mathrm{BB}$, for the total standard sample and each spectroscopic type. Uncertainties are the 16th--84th percentiles of the bootstrap (\citealt{Gehrels1986} limits for $N \leq 2$). Bins in which a type has no events are marked with an ellipsis.}
\label{tab:lf_bb}
\renewcommand{\arraystretch}{1.7}
\setlength{\tabcolsep}{3pt}
\begin{tabular}{l | c c c c c}
\toprule
 & & \multicolumn{4}{c}{$\log_{10}(\phi\,[\Runit\,\mathrm{dex^{-1}}])$} \\
\cmidrule(lr){3-6}
$\log_{10}(L_{\rm BB}/\mathrm{erg\,s^{-1}})$ & $N$ & Total & TDE-H & TDE-He & TDE-H+He \\
\midrule
42.5--43.1 & 1 & $1.98^{+0.52}_{-0.76}$ & \nodata & \nodata & $1.98^{+0.52}_{-0.76}$ \\
43.1--43.7 & 2 & $1.55^{+0.37}_{-0.45}$ & \nodata & \nodata & $1.55^{+0.37}_{-0.45}$ \\
43.7--44.3 & 11 & $1.92^{+0.15}_{-0.23}$ & $1.00^{+0.20}_{-0.40}$ & $0.08^{+0.52}_{-0.76}$ & $1.86^{+0.17}_{-0.28}$ \\
44.3--44.9 & 6 & $2.13^{+0.22}_{-0.55}$ & \nodata & \nodata & $2.13^{+0.22}_{-0.55}$ \\
44.9--45.5 & 1 & $-0.13^{+0.52}_{-0.76}$ & \nodata & $-0.13^{+0.52}_{-0.76}$ & \nodata \\
\bottomrule
\end{tabular}
\end{table*}

\begin{table*}
\centering
\caption{TDE rates and rate enhancement factors $\Gamma$ (Equation~\ref{eq:enhancement}) in every H$\alpha$ EW--Lick H$\delta_A$ region used in Section~\ref{subsec:host_qbs_psb}, relative to the stellar-mass- and redshift-matched SDSS control sample. Uncertainties are the 16th--84th percentiles of the joint Poisson bootstrap.}
\label{tab:enhancement_all}
\renewcommand{\arraystretch}{1.7}
\setlength{\tabcolsep}{2.8pt}
\begin{tabular}{l | c c c c c c c c}
\toprule
Region & H$\alpha$ EW [\AA] & Lick H$\delta_A$ [\AA] & $N_\mathrm{TDE}$ & $f_\mathrm{rate,TDE}$ & $f_\mathrm{control}$ & $\Gamma$ & $\Gamma_\mathrm{excl}$ & $\log_{10}(R\,[\Runit])$ \\
\midrule
PSB & $<3$ & $\geq 4$ & 4 & $0.070^{+0.070}_{-0.051}$ & $0.0046^{+0.0017}_{-0.0015}$ & $15^{+16}_{-11}$ & $27^{+18}_{-18}$ & $1.16^{+0.23}_{-0.57}$ \\
QBS & $<3$ & $1.3$--$4$ & 3 & $0.081^{+0.083}_{-0.067}$ & $0.018^{+0.004}_{-0.003}$ & $5^{+5}_{-4}$ & $6^{+6}_{-5}$ & $1.23^{+0.25}_{-0.72}$ \\
PSB+QBS & $<3$ & $\geq 1.3$ & 7 & $0.15^{+0.12}_{-0.08}$ & $0.023^{+0.006}_{-0.004}$ & $7^{+5}_{-3}$ & $10^{+6}_{-5}$ & $1.50^{+0.19}_{-0.32}$ \\
\midrule
\multicolumn{9}{l}{\textit{Coarse grid (Figures~\ref{fig:Halpha_Hdelta_rates} and \ref{fig:Halpha_Hdelta_enhancement}, left panels)}} \\
 & $-2$--$10.3$ & $-4$--$0$ & 4 & $0.11^{+0.11}_{-0.09}$ & $0.34^{+0.15}_{-0.16}$ & $0.32^{+0.61}_{-0.28}$ & $0.33^{+0.64}_{-0.29}$ & $1.35^{+0.27}_{-0.93}$ \\
 & $-2$--$10.3$ & $0$--$4$ & 7 (6) & $0.51^{+0.21}_{-0.22}$ & $0.12^{+0.03}_{-0.02}$ & $4.2^{+1.9}_{-2.1}$ & $2.3^{+0.9}_{-0.9}$ & $2.03^{+0.20}_{-0.39}$ \\
 & $-2$--$10.3$ & $4$--$8$ & 4 & $0.070^{+0.070}_{-0.051}$ & $0.014^{+0.006}_{-0.006}$ & $5.1^{+8.6}_{-3.6}$ & $12^{+8}_{-8}$ & $1.16^{+0.23}_{-0.57}$ \\
 & $10.3$--$22.7$ & $-4$--$0$ & 1 & $0.022^{+0.050}_{-0.018}$ & $0.0058^{+0.0025}_{-0.0025}$ & $3.7^{+8.6}_{-3.1}$ & $3.7^{+8.6}_{-3.1}$ & $0.65^{+0.52}_{-0.76}$ \\
 & $10.3$--$22.7$ & $0$--$4$ & 1 & $0.0017^{+0.0040}_{-0.0014}$ & $0.094^{+0.025}_{-0.014}$ & $0.019^{+0.043}_{-0.015}$ & $0.022^{+0.051}_{-0.018}$ & $-0.44^{+0.52}_{-0.76}$ \\
 & $10.3$--$22.7$ & $4$--$8$ & 1 & $0.26^{+0.59}_{-0.21}$ & $0.048^{+0.021}_{-0.027}$ & $5.3^{+12.2}_{-4.4}$ & $16^{+36}_{-13}$ & $1.73^{+0.52}_{-0.76}$ \\
 & $22.7$--$35$ & $-4$--$0$ & 0 & \nodata & $0.0025^{+0.0011}_{-0.0010}$ & \nodata & \nodata & \nodata \\
 & $22.7$--$35$ & $0$--$4$ & 1 & $0.023^{+0.053}_{-0.019}$ & $0.067^{+0.018}_{-0.012}$ & $0.35^{+0.80}_{-0.29}$ & $0.49^{+1.12}_{-0.40}$ & $0.68^{+0.52}_{-0.76}$ \\
 & $22.7$--$35$ & $4$--$8$ & 1 & $0.0040^{+0.0091}_{-0.0033}$ & $0.052^{+0.021}_{-0.025}$ & $0.076^{+0.174}_{-0.063}$ & $0.18^{+0.42}_{-0.15}$ & $-0.08^{+0.52}_{-0.76}$ \\
\bottomrule
\end{tabular}

\vspace{4pt}
\begin{minipage}{\textwidth}
\raggedright
\textit{Note.} The PSB, QBS, and PSB+QBS rows repeat Table~\ref{tab:tde_enhancement} with the TDE rate added. $\Gamma_\mathrm{excl}$ excludes TDE 2023clx, whose host carries over a quarter of the total sample weight; where this changes $N_\mathrm{TDE}$, the reduced number is given in parentheses. Grid cells include their lower edges and exclude their upper edges. For regions with $N_\mathrm{TDE} \leq 2$, the uncertainties on $f_\mathrm{rate,TDE}$, $\Gamma$, and $R$ are the \citet{Gehrels1986} $1\sigma$ limits scaled to the measured value (Section~\ref{subsec:vol_rates}); $f_\mathrm{control}$ uses the bootstrap throughout. All 20 TDE hosts with host-galaxy fits lie within the grid, but 26\% of the control weight lies outside the plotted ranges, so $f_\mathrm{control}$ sums to 0.74 over the grid cells. $f_\mathrm{rate,TDE}$ and $R$ are for the full sample.
\end{minipage}
\end{table*}

\end{appendices}

\end{document}